\documentclass[12pt,preprint]{aastex}

\slugcomment{Accepted for Publication in \apj, March 22, 2011}
\shorttitle{Polarized Radio Sources}
\shortauthors{Banfield et al.}

\begin{document}
\title{Polarized Radio Sources: A Study of Luminosity, Redshift  and Infrared Colors}
\author{Julie K.~Banfield\footnote{Current address: CSIRO Australia Telescope National Facility, PO Box 76, Epping, NSW, 1710, Australia}, Samuel J.~George, A.~Russ~Taylor, Jeroen M.~Stil}
\affil{Institute for Space Imaging Science, Department of Physics and Astronomy, \\ University of Calgary, Alberta, T2N 1N4, Canada}
\author{Roland Kothes}
\affil{Dominion Radio Astrophysical Observatory, Herzberg Institute of Astrophysics, \\ National Research Council Canada, Penticton, BC V2A 6J9, Canada}
\author{Douglas Scott}
\affil{Department of Physics and Astronomy, University of British Columbia, Vancouver, BC V6T 1Z1, Canada}

\begin{abstract}
The Dominion Radio Astrophysical Observatory Deep Field polarization study has been matched with the \textit{Spitzer} Wide-Area Infrared Extragalactic survey of the European Large Area \textit{Infrared Space Observatory} Survey North 1 field.  We have used VLA observations with a total intensity rms of $87\,\mu$Jy beam$^{-1}$ to match SWIRE counterparts to the radio sources.  Infrared color analysis of our radio sample shows that the majority of polarized sources are elliptical galaxies with an embedded active galactic nucleus.  Using available redshift catalogs, we found 429 radio sources of which 69 are polarized with redshifts in the range of $0.04<z<3.2$.  We find no correlation between redshift and percentage polarization for our sample.  However, for polarized radio sources, we find a weak correlation between increasing percentage polarization and decreasing luminosity.   
\end{abstract}


\section{Introduction}
Cosmic magnetic fields have become of increasing importance in understanding the Universe.  Magnetic fields play an essential role in star formation and are believed to be part of galaxy formation and evolution.  However, little is known about cosmic magnetic field origin, structure, and evolution.  Do magnetic fields and galaxies evolve together?  How are large scale magnetic fields generated and maintained?  What is the role of magnetic fields in galaxy evolution?  Answers to these questions have become one of the five main science goals of the next generation radio telescope, The Square Kilometer Array \citep[SKA,][]{Schillizzi2004}.

Detecting radio waves provides information on the magnetic field structure of an object.  Extragalactic radio sources emit synchrotron radiation which is polarized, where the percentage polarization theoretically can be as large as $70\,$\% \citep{Pacholczyk1970}.  Synchrotron emission provides the measure of the total field strength and the polarization of the radio wave provides information on the direction of the magnetic field and the degree of ordering.  Information on the relationship between the environments of the radio sources and their properties can be inferred by the magnetic field structure.  Understanding of magnetic fields can lead to studies of the evolution of magnetic fields and the environments of radio sources with redshift and luminosity.  

At $1.4\,$GHz, active galactic nuclei (AGNs) are detected over a luminosity range between $10^{22} - 10^{28}\,$W Hz$^{-1}$ and out to redshifts $z > 1$.  The majority of extragalactic polarized sources that have been studied are radio-loud AGN.  AGNs can be separated into a mixture of extragalactic radio sources which include Fanaroff-Riley class I and II radio galaxies \citep{FR1974}, where the classification is based on morphology and luminosity.  \citet{Jackson1999} show a model for an AGN where an orientation close to the line-of-sight will result in a quasar or blazar while the classic FRII is observed when the orientation angle is perpendicular to the line-of-sight.   FRIIs dominate the higher luminosity range and are more powerful than FRIs, however, the boundary divide between FRIIs and FRIs is unknown as there is considerable overlap between the two classes in the luminosity function \citep{Jackson2004}.  The classical FRII has highly collimated jets originating from the core with the large lobe structure getting brighter as they extend away from the core, with hotspots near the outer edge of the lobe.  FRIs, on the other hand, have brighter emission near the core and decreases further away.  The magnetic fields in FRIIs are shown to line up along the jet axes with the majority of the polarized emission originating in the hotspots.  On the other hand, in FRIs the magnetic fields lie perpendicular to the source axes and are shown to wrap around the jet \citep[see][and references within]{Saikia1988}.   

Recent studies of polarized radio sources have shown that percentage polarization increases with decreasing flux density \citep{Mesa2002,Tucci2004,Sadler2006,Taylor2007,Grant2010,atlbs2010}.  \citet{Mesa2002} suggest that a population change of radio sources at fainter flux densities causes the increasing percentage polarization with decreasing flux density.  \citet{Taylor2007} suggest that the changing fraction of radio-quiet AGN is responsible for the increase in polarization at faint flux density levels.  \citet{atlbs2010} propose that the increase in percentage polarization with decreasing flux is the result of the transition from an FRII-dominated population to an FRI-dominated population.  The cause of this increase in $\Pi$ with decreasing flux density remains unknown.

Polarized observations of resolved and compact radio sources have been investigated by \citet{Mesa2002} and \citet{Grant2010} to study the increase in $\Pi$ with decreasing flux density.  For radio sources with $S_{\rm 1.4} \ge 500 \,$mJy, \citet{Mesa2002} determined that compact polarized radio sources (unresolved at $45\arcsec$) are more polarized than resolved polarized sources.  On the other hand, for radio sources with $S_{\rm 1.4} < 500\,$mJy, \citet{Grant2010} found that sources resolved at $5\arcsec$ are more polarized than compact sources.  \citet{Grant2010} suggest that the higher degree of polarization may be originating from the unbeamed lobe-dominated structure, not beamed blazars.  \citet{Shi2010} study a sample of highly polarized objects to determine the nature of these radio sources.  \citet{Shi2010} show that polarized radio sources with percentage polarization $\Pi>30\%$ are contained in elliptical galaxies and that the sample of highly polarized sources have luminosity densities in the range of $L_{\rm 1.4}=10^{23}-10^{24}$ W Hz$^{-1}$.  However, \citet{Shi2010} found no difference in the source environments between low polarization sources and ultrahigh polarization sources, indicating that the high polarization must be a result of intrinsic properties of the radio sources and not dependent on optical morphology, redshift, linear size, and radio power.

We present an analysis of $1.4\,$GHz polarization data with the \textit{Spitzer} Wide-Area Infrared Extragalactic Survey \citep[SWIRE,][]{Lonsdale2003} observations to study the nature of polarized radio sources to investigate possible reasons for the observed increase in $\Pi$ with decreasing flux density.  We use both spectroscopic and photometric redshifts presented by \citet{RR2008} and the Sloan Digital Sky Survey Data Release 7 \citep[SDSS DR7,][]{sdssdr7}.  We outline the sample selection and data processing in section \ref{obs}.  Section \ref{cat} describes the polarized source list containing $1.4\,$GHz radio parameters and the SWIRE infrared source properties.  The nature of our polarized source sample is described in section \ref{nature}. The cosmological parameters used throughout this paper are: $\Omega_\Lambda=0.7$, $\Omega_{\rm M}=0.3$, and $H_0=70\,{\rm km\,s^{-1}\,Mpc^{-1}}$.   Spectral index is defined as $S\sim\nu^{\alpha}$ between $1.4\,$GHz and $325\,$MHz.

\section{Sample Selection and Observations}\label{obs}
The data presented in this paper cover the European Large Area \textit{ISO} Survey North 1 region \citep[ELAIS N1,][]{Oliver2000} centered on $\alpha_{\rm 2000} = 16^{\rm h}10^{\rm m}$ and $\delta_{\rm 2000} = 54\arcdeg30\arcmin36\arcsec$.  Table \ref{enobs} provides a listing of available data in the ELAIS N1 region.

\subsection{$1.4\,$GHz observations and data}
The 1.4 GHz radio source sample used in this paper was taken from \citet{Grant2010} and was observed with the Dominion Radio Astrophysical Observatory synthesis telescope \citep[DRAO ST,][]{Landecker2000}.   The observations achieved a sensitivity of $55\,\mu$Jy beam$^{-1}$ in total intensity and $45\,\mu$Jy beam$^{-1}$ in both Stokes $Q$ and $U$.  The survey covers 15.16 deg$^2$, centered on $\alpha_{\rm 2000}=16^{\rm h}14^{\rm m}$ and $\delta_{\rm 2000}=54\arcdeg56\arcmin$, and contains 958 radio sources down to $S_{1.4} = 440\,\mu$Jy beam$^{-1}$ ($8.0\,\sigma_{\rm I}$), of which 136 are detected in polarization down to bias-corrected $p_{\rm 0} = 261\,\mu$Jy beam$^{-1}$ ($5.8\,\sigma_{\rm QU}$).  The angular resolution is $42\arcsec \times 69\arcsec$ at the field center.  Instrumental polarization levels range from $0.5\,\%$ at the field center to $1\,\%$ at the edges of the mosaic.

\begin{figure}
\begin{center}
\includegraphics[scale=0.8]{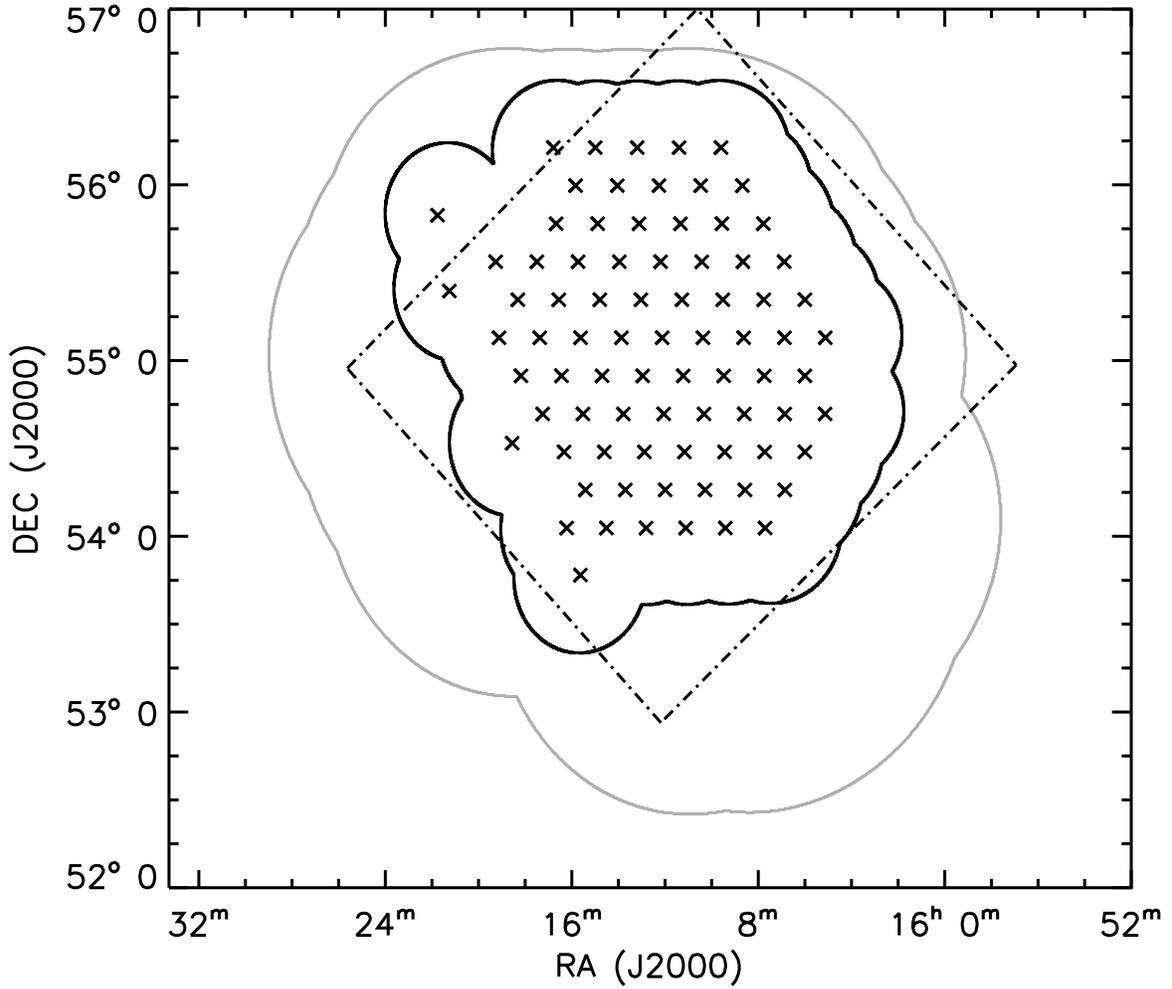}
\caption{The VLA observations (solid black line) covering the area of the DRAO deep field region (solid grey line).  The crosses show the location of the 80 individual pointing centers, each separated by $15\arcmin$.  The dot-dashed box outlines the SWIRE area.}
\label{obsarea}
\end{center}
\end{figure}

To obtain cross-identifications with SWIRE counterparts, higher resolution imaging is required to provide accurate positions.  High resolution ($\theta=3.9\arcsec$) observations of the DRAO \textit{Planck} Deep Fields ELAIS N1 region were carried out with the VLA in 2007 October.  The observations were made by the VLA in B configuration for seven $3.125\,$MHz channels centered on $1.465\,$GHz.   The observations lasted a total of $31.5\,$hours, split into three days of $10.5\,$hours.  The area was covered by 80 pointings, each separated by $15\,\arcmin$ in a hexagonal pattern and centered at $\alpha_{2000}=16^{{\rm h}}13^{{\rm m}}$ and $\delta_{2000}=55\arcdeg$. Fig.$\,\,$\ref{obsarea} shows the VLA mosaic geometry with respect to the DRAO deep field observations and the SWIRE ELAIS N1 area.

The image processing was completed within the Astronomical Image Processing System (AIPS) data reduction package.  The three days were calibrated and edited separately using the polarized flux calibrator 3C286 and the phase calibrator 1634+627.  3C286 was observed twice each day while the phase calibrator was observed every 20 to 30 minutes.  The flux calibration used the Baars flux density scale \citep{Baars1977} of $S_{\rm 1.4}=14.8\,$Jy for 3C286.  At the time of the observations, 12 of the antennas had been upgraded to Expanded VLA (EVLA) specifications.  The amplitudes of the EVLA antennas on the EVLA-VLA baselines were found to occasionally drop on the third day.   As a result, all EVLA-VLA baselines were flagged on the third day and otherwise the EVLA antennas remained in the calibration and imaging observations.

All 80 pointings were channel averaged and imaged out to $80\%$ of the primary beam as the mapping efficiency does not improve beyond this point \citep[see][]{Condon1998}.  To minimize geometric phase distortions, each pointing was split into 61 facets to cover the primary beam area, each with $500\times500$ pixels.  All facets were {\tt CLEAN}ed and self-calibrated and then combined into one image using the AIPS task {\tt FLATN}.  {\tt FLATN} was then run a second time to mosaic all 80 pointings together and apply the primary beam correction.  The resulting total intensity rms noise level at the center of the mosaic is $87\,\mu$Jy beam$^{-1}$.

As shown in Fig.$\,\,$\ref{obsarea}, the VLA observations do not completely cover the DRAO deep field or the SWIRE ELAIS N1 area.  For the region outside the the VLA mosaic, the Faint Images of the Radio Sky at Twenty Centimeters \citep[FIRST,][]{White1997} survey was used to provide high resolution ($\theta=5\arcsec$) imaging of those DRAO deep field radio sources down to $S_{\rm 1.4}=2.0\,$mJy  ($5\sigma$).  Classification of each radio galaxy in the study was completed by \citet{Grant2010} who made the distinction between compact and resolved structure using FIRST at an angular resolution of $5\arcsec$.  This classification is used in our analysis.  The polarization information for the radio sources comes from the DRAO deep field source catalog \citep{Grant2010}.  

\subsection{\textit{Spitzer} Infrared Data}
The infrared and optical observations of the $8.72$ deg$^{2}$ ELAIS N1 region were taken from the SWIRE ELAIS N1 source catalog \citep{Surace2004}.  There are 662 DRAO deep field radio sources covering the SWIRE ELAIS N1 field, of which 553 radio sources have an accurate position to match to SWIRE objects from the VLA observations and the FIRST positions, while 109 have only the DRAO position.  The 109 radio sources with only a DRAO position were covered by either the VLA observations or FIRST but were not detected because: (1) the flux density is below the detection threshold of the VLA observations and FIRST; or (2) the diffuse emission is resolved out by the higher resolution imaging. 

Cross-identifications between SWIRE and VLA/FIRST positions were made using a search radius of $3\arcsec$ around the high resolution VLA radio position and taking the closest SWIRE source.  The mean positional offset between the radio source position and the SWIRE position is $-0.15\arcsec$ in right ascension and $0.12\arcsec$ in declination.  The probability of a false cross-identification between the radio source and a SWIRE source was determined by shifting the radio positions by $1\arcmin$ in four directions (up, down, left, and right).  For each shift in position, the cross-identification procedure was repeated.  The mean probability of a false cross-identification was found to be $5.2\,\%$ for a search radius of $3\arcsec$.  

Visual inspection of all 662 radio sources in the SWIRE ELAIS N1 images were completed in order to confirm the matches.  An additional complication is that a resolved radio source can have multiple components which can lead to misidentification with a SWIRE source.  Visual inspection of multiple component radio sources avoided this problem.  The 662 DRAO ELAIS N1 radio sources in the SWIRE ELAIS N1 images were separated into four categories, see Table \ref{swireid}:
\begin{enumerate}
\item Unambiguous - Exact matches to a SWIRE ELAIS N1 source, of which we found 444 radio sources.  Exact matches were made using only those radio sources with high resolution VLA or FIRST positioning.
\item Ambiguous - There is no high-resolution position (a result of resolution and sensitivity) to help determine the position of the SWIRE counterpart or the radio source is extended so that the central galaxy could not be positively identified.  There are 166 objects in this category.
\item Infrared Faint - Radio sources with high resolution imaging but no SWIRE ELAIS N1 counterpart. We found 18 radio sources matching this criteria.  These objects could possibly be infrared faint radio sources \citep[IFRS,][]{Norris2006}.
\item Not Cataloged - There are 34 radio sources with a \textit{Spitzer} counterpart that did not meet the SWIRE ELAIS N1 catalog flux density cutoff and were therefore not included in the SWIRE ELAIS N1 catalog.
\end{enumerate}
Only the radio sources with an unambiguous match in the SWIRE catalog were used for the subsequent analysis.  These sources represent $67\,\%$ of our radio source sample.  

\subsection{Available Redshift catalogs}
Photometric redshifts for the DRAO deep field sources were taken from the \citet{RR2008} photometric redshift catalog (hereafter RR08) of the SWIRE survey.  RR08 have cataloged photometric redshifts for 218,117 ELAIS N1 sources.  Using the 444 radio sources with an exact SWIRE identification, we obtained photometric redshifts for 189 radio sources in RR08.  The best fit photometric redshift was used in order to minimize $\chi^2$, resulting in using the $A_V=0$ solution in all cases.  The difference between the photometric redshifts fitted with the $A_V=0$ template and the freely fitted $A_V$ template was negligible. We also required our photometric redshifts to have: (1) a fit using four or more photometric bands; and (2) a reduced $\chi^2< 10$.  These limits are similar to those RR08 used to compare photometric redshifts with spectroscopic redshifts and RR08 noted that a $\chi^2 > 10$ is a failure of the photometric method.  

\begin{figure}
\begin{center}
\includegraphics[scale=0.8]{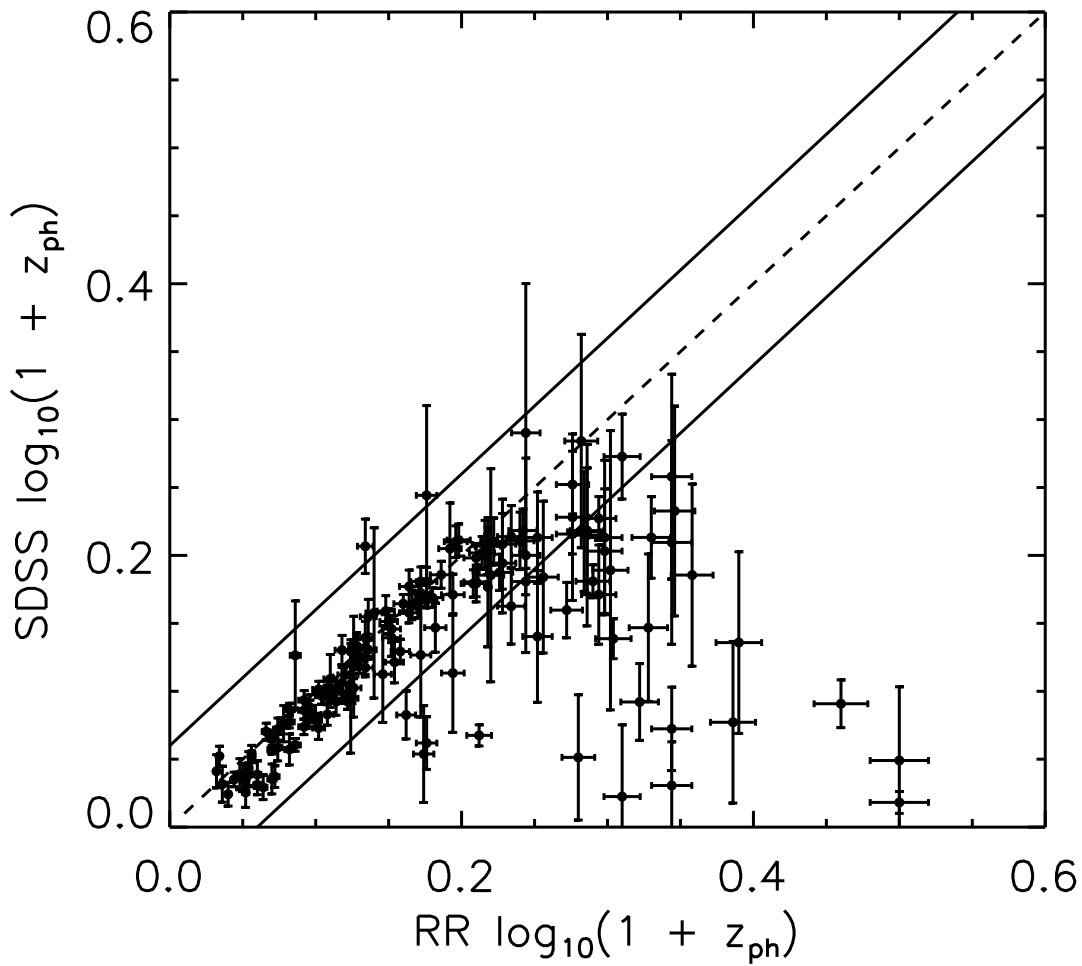}
\caption{Comparison of 150 photometric redshifts from \citet{RR2008} and SDSS DR7 for DRAO ELAIS Deep Field radio sources.  The diagonal dashed line is $y=x$, while the solid lines indicate $|$log$_{10}(1+z_{\rm SDSS})/(1+z_{\rm ph})| \le 0.06$.} \label{sdssplot}
\end{center}
\end{figure}

In order to test the RR08 photometric redshifts, we matched photometric redshifts of SDSS DR7 sources for 154 radio sources that also have a photometric redshift in RR08.  Fig.$\,\,$\ref{sdssplot} shows the comparison of the photometric redshifts between RR08 and SDSS DR7 for radio sources whose position is within $5\arcsec$ in both catalogs.  The $|$log$_{10}(1+z_{\rm SDSS})/(1+z_{\rm ph})|\le 0.06$ restriction from RR08 is also shown on the plot and the majority of our radio sources lie within this cutoff.  The outliers are predominantly found at $z>1$, where the SDSS DR7 errors are large. The errors in the RR08 photometric redshifts depend on the number of photometric bands available for the fit, the limiting magnitude of the object, and the limiting value of $\chi^2$.  RR08 determined that the typical rms value is as low as $2.5\,\%$ when fitted with 13 photometric bands and as large as $6\,\%$ when fitted with four photometric bands.  For QSOs fitted with four or more photometric bands, the uncertainty can be as large as $10\,\%$.  Therefore the photometric redshift uncertainty in the RR08 catalog is very wide and depends heavily on the type of object and the number of photometric bands included in the fit.  It should be noted that RR08 noticed a slight systematic overestimation around $z \sim 1$ by 0.1 in ELAIS N1 that was not seen in the other SWIRE fields.  RR08's explanation of this overestimation is the result of a bias in the photometry at fainter magnitudes.

For radio sources outside the SWIRE ELAIS N1 region, photometric redshifts were obtained from the SDSS DR7 for sources with high resolution VLA imaging.  If a radio source has a photometric redshift in both RR08 and SDSS DR7, the RR08 photometric redshift was used.  We used spectroscopic redshifts instead of photometric redshifts when available.  Spectroscopic redshifts were obtained for 24 radio sources from RR08 as well as the SDSS DR7.   In our DRAO deep field radio source list we were able to identify one radio source in the \citet{Swinbank2007} supercluster; DSX1-25.  The corresponding radio source has no detectable polarization and has a $1.4\,$GHz flux density of $0.77 \pm 0.12\,$ mJy, $\alpha^{1420}_{325}>-2.13$, and $L_{\rm 1.4} \sim 3 \times 10^{24}\,$W Hz$^{-1}$ .  In summary, we have obtained redshift information for 429 radio sources, of which 69 are polarized in the DRAO deep field.

\begin{figure}
\includegraphics[scale=0.8]{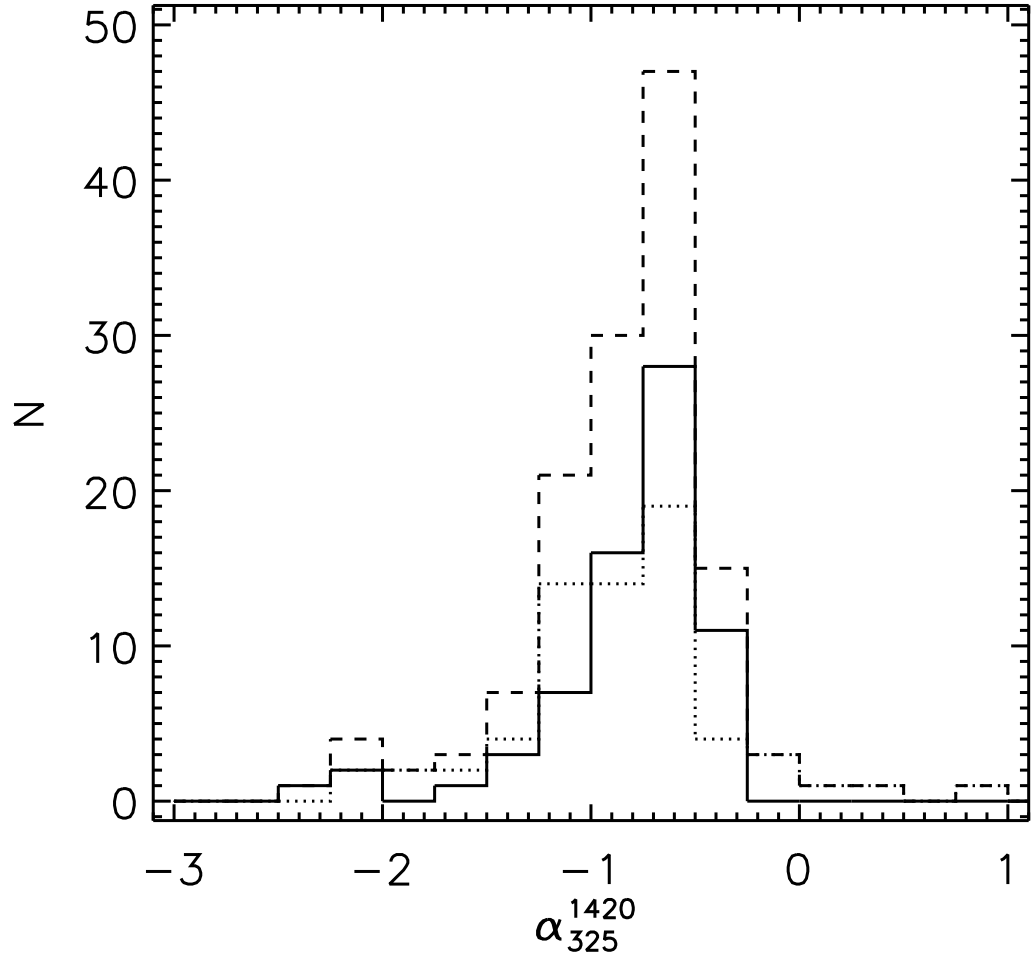}
\caption{Spectral index of all 136 polarized sources (dashed line) in the DRAO deep field.  Polarized sources with known redshift and unknown redshift are indicated by the solid and dotted lines, respectively.} \label{alphahist}
\end{figure}

\begin{figure}
\includegraphics[scale=0.8]{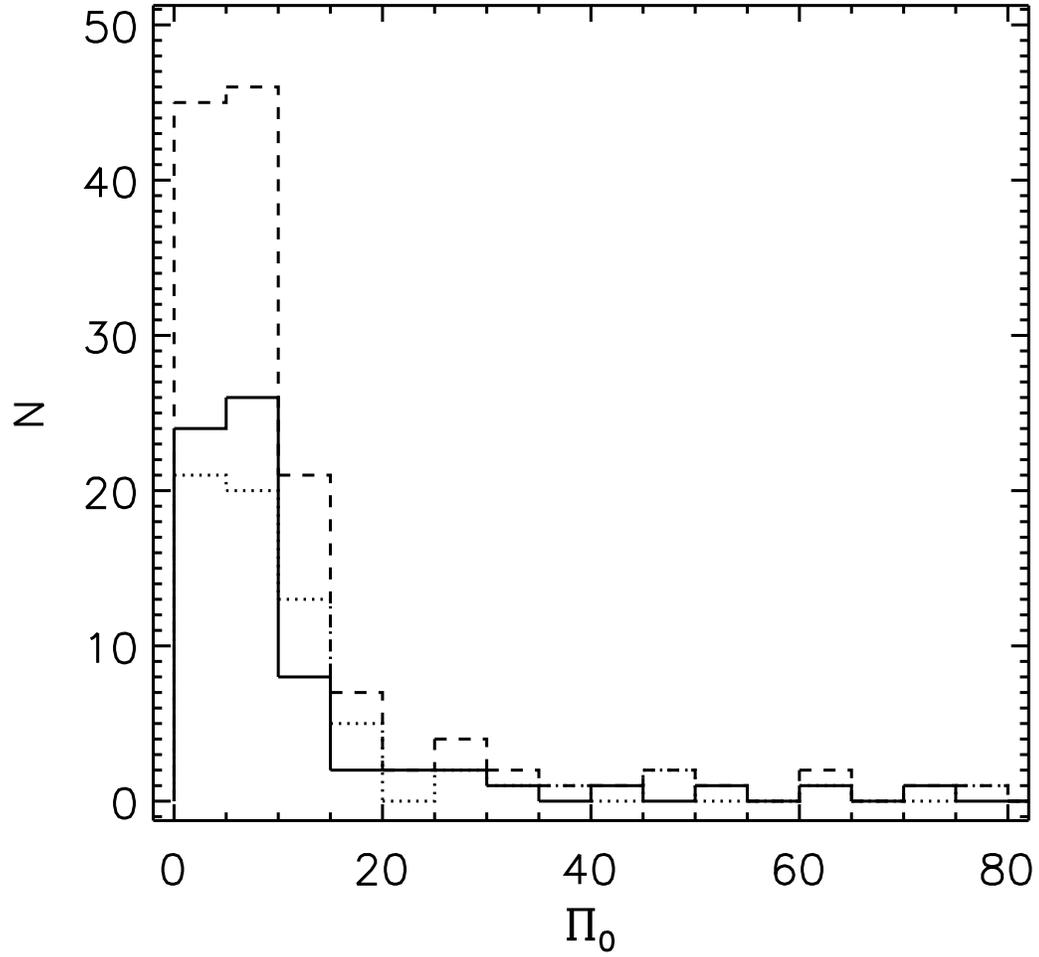}
\caption{Percentage polarization for DRAO deep field polarized sources (dashed line). Polarized sources with known redshift and unknown redshift are indicated by the solid and dotted lines, respectively.} \label{pihist}
\end{figure}

To test the completeness of our radio sources with redshifts, we compared the spectral index and percentage polarization distribution of polarized radio sources with and without redshifts, shown in Fig.$\,\,$ \ref{alphahist} and \ref{pihist}.  In both cases the sample of polarized sources with a redshift is similar to those with no redshift.  For the spectral index distribution of polarized sources with and without a redshift, the two-dimensional KS test determined a significance level of 0.19.  The significance level ranges form 0 to 1 and a small value indicates that the cumulative distribution function of the two data sets is significantly different.  For the percentage polarization distribution of polarized sources with and without a redshift, the two-dimensional KS test determined a significance level of 0.34.

%%% SOURCE catalog %%%
\section{Merged Source catalog}\label{cat}
The luminosity density of the 429 radio sources with redshifts was calculated using the equations from \citet{Hogg1999} and including the $k$-correction:
\begin{equation}
L_{\rm 1.4} = \frac{4\pi D^2_{\rm L} S_{\rm 1.4}}{(1+z)}(1+z)^{-\alpha} \ ,
\end{equation}
where $D_{\rm L}$ is the luminosity distance, $S_{\rm 1.4}$ is the flux density at $1.4\,$GHz, $\alpha$ is the spectral index ($S\sim \nu^{\alpha}$) from \citet{Grant2010} and $z$ is the redshift.  Upper limits for the spectral index were only available for 18 of the 69 polarized sources and 270 of the 360 radio sources with no detectable polarization.  For sources with only an upper limit to the spectral index, a value of $\alpha=-0.6$ for compact objects and $\alpha=-0.9$ for resolved objects was assumed, based on the mean value for each classification in \citet{Grant2010}.  Linear polarization $p$ is evaluated using Stokes $Q$ and Stokes $U$ where $p\equiv\sqrt{Q^2+U^2}$.  The estimate of the bias-corrected polarized flux is $p_{\rm 0} = \sqrt{p^2 - \sigma^2}$ for a signal-to-noise ratio greater than 4 \citep{Simmons1985}.  The percentage polarization is evaluated by:
\begin{equation}
\Pi \equiv \frac{p}{S} \times 100\,\% \ ,
\end{equation}
where S is the Stokes $I$ total flux density of the source.   The bias-corrected percentage polarization $\Pi_{\rm 0}$ is determined using $p_{\rm o}$ instead of $p$.  The 69 polarized radio sources with a redshift are listed in Table \ref{catalog}, which contains the DRAO deep field name and $1.4\,$GHz properties from \citet{Grant2010}, as well as the redshift and luminosity. 

\section{The Nature of Polarized Radio Sources}\label{nature}
\subsection{\textit{Spitzer} Infrared Classification}\label{swire}
\begin{figure}
\begin{center}
\includegraphics[scale=0.8]{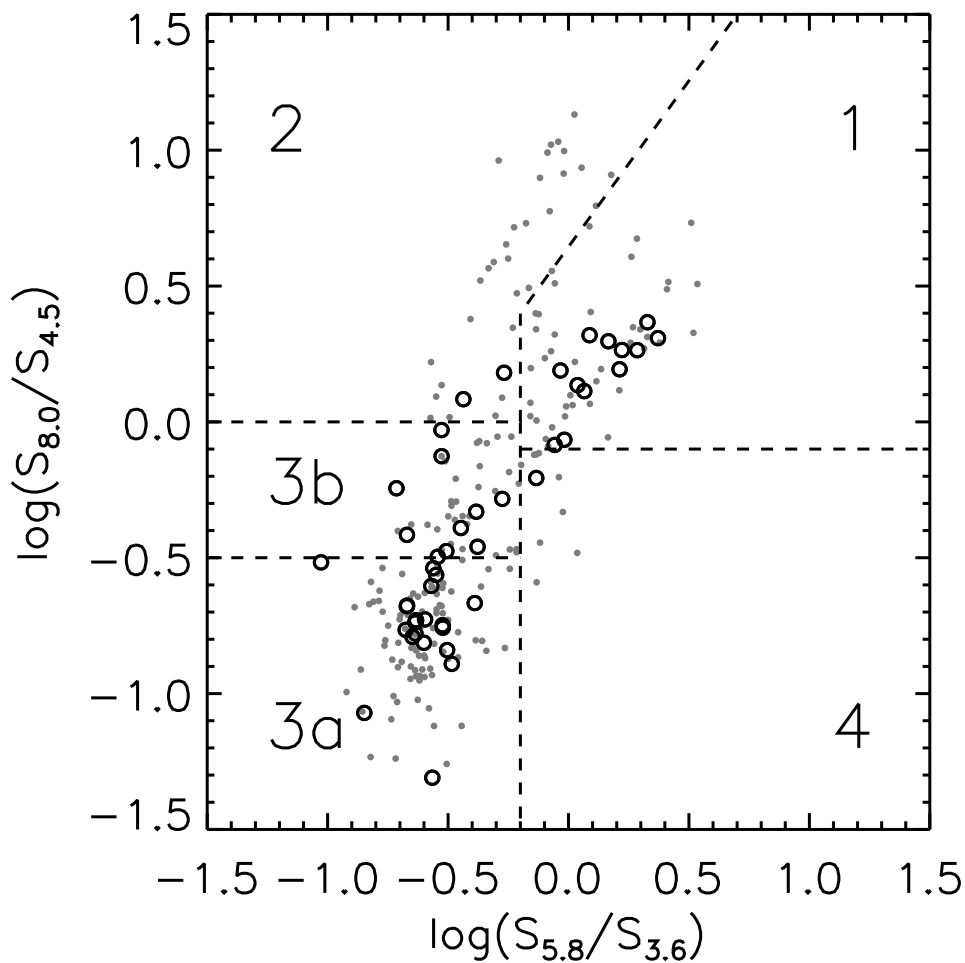}
\caption{\textit{Spitzer} near-infrared color-color diagram of the DRAO deep field host galaxies which have a SWIRE detection in all four near-infrared bands.  Open black circles indicate polarized radio sources, while the grey closed circles indicate the sources with no detectable polarization.  The five regions outlined by the dashed lines are based on the division by \citet{Sajina2005} and are outlined in the text.} \label{swirecc}
\end{center}
\end{figure}

The near-infrared bands of the SWIRE ELAIS N1 catalog provide insight into the type of galaxy that hosts the radio emission.  The $3.6-8.0\,\mu$m infrared wavelength range can distinguish emission between stellar light in the host galaxy and light that is reprocessed by dust and gas \citep{Sajina2005}.  Fig.$\,\,$\ref{swirecc} shows the \textit{Spitzer} near-infrared color-color diagram of 243 DRAO deep field host galaxies that were detected in all four near-infrared \textit{Spitzer} bands: 3.6, 4.5, 5.8, and 8.0$\,\mu$m.  There are 45 polarized radio sources and 198 sources with no detectable polarization matching these criteria in our sample.  The majority of our sources are identified with extended infrared galaxies for which we have selected the SWIRE isophotal integrated fluxes for the four bands in our analysis.  The color-color diagram is split into five regions based on the simulations of \citet{Sajina2005}.  The divisions are determined by the strength of the polycyclic aromatic hydrocarbon (PAH) bands and the slope of the continuum in various redshift ranges.  Region 1 preferentially selects sources where the infrared emission is dominated by non-equilibrium emission of very small dust grains, which is interpreted as PAH destruction by the hard ultraviolet spectrum of an AGN.  Region 2 is mainly populated by dusty star-forming galaxies at redshift $z<0.5$ with strong PAH bands, as the 3.6 and 8.0$\,\mu$m flux contain the strongest PAH features at low redshift.  Region 3 is populated by: (a) elliptical galaxies dominated by the starlight of old stellar populations; and (b) galaxies with fainter PAH emission.  This region can be split into the proposed galaxy types by the lines log$(S_{8.0}/S_{4.5})\le-0.5$ for region 3a and log$(S_{8.0}/S_{4.5}) >-0.5$ for region 3b as denoted by \citet{Sajina2005} and \citet{Taylor2007}.  Region 4 consists of PAH-dominated galaxies at redshift $z=1.5-2$.  The number of sources found in each SWIRE color-color region is listed in Table \ref{swirenum}.

The polarized radio sources lie mainly in regions 1 and 3, with $67\%$ of the polarized sample in region 3.  Therefore the majority of our \textit{polarized} radio source population tend to have host elliptical galaxies, confirming results by \citet{Taylor2007}.  The polarized sources in region 1 tend to be infrared galaxies with very small dust grains where the radio emission indicates an AGN.  Our color-color diagram distribution differs from \citet{Taylor2007} in region 3.  There are polarized sources in region 3 of the earlier diagram that are not included in our diagram.  We have adapted a more conservative signal-to-noise polarization cutoff than \citet{Taylor2007} which has removed some of these sources in both total intensity and polarization.  \citet{Taylor2007} noted that there was an excess of polarized radio sources in region 3a compared to radio sources with no detectable polarization.  The higher flux density cutoff of our sample does not confirm the earlier result, indicating that the results of \citet{Taylor2007} could come from a dependence on the flux density detection level of radio sources.  The additional radio sources in region 3a that were detected  are not sufficiently bright to enable us to detect the polarized emission.  A more sensitive polarization survey should detect the polarization from these fainter radio sources.

In our radio source sample, the radio sources that lie in region 2 are radio sources with infrared emission dominated by PAH indicating a star-forming galaxy. Table \ref{spiral} lists the 16 spiral galaxies in our sample with upper limits on $\Pi_{\rm 0}$ for each spiral galaxy using an upper limit of $p_{\rm UL} = p + 2\sigma_{\rm QU}$ from \citet{Vaillancourt2006}.  \citet{Stil2009} show that spiral galaxies detected at $4.8\,$GHz can have $\Pi_{\rm 0}$ as large as $18\,$\% indicating that the upper limits on $\Pi_{\rm 0}$ for our spiral galaxies are larger and a deeper polarization survey should detect polarized emission from the majority of these spiral galaxies.  However, there are face-on spiral galaxies with clearly visible spiral arms that have small $\Pi_{\rm 0}$ in our list.  Beam depolarization and not sensitivity could be the reason why these face-on spiral galaxies have no detectable polarization.

\subsection{Infrared Faint Radio Sources}\label{ifrs}
Most radio sources have detectable infrared counterparts.  Sources without infrared counterparts, the so-called infrared faint radio sources (IFRS) were first discovered by \citet{Norris2006}.  \citet{Norris2006} provide four possible explanations for what these radio sources are: (1) AGN or star-forming galaxy that have a high redshift or are very heavily obscured; (2) AGN or starburst in a transitory phase; (3) unidentified radio lobe; or (4) unknown object, potentially Galactic in origin.  Observations of IFRS have suggested that these objects are high redshift AGN of the 3C type at $z > 2$ \citep{Garn2008b} or compact steep spectrum sources with $z>1$ \citep{Middelberg2008}.  Recent observations by \citet{Norris2011} suggest that IFRS are dust-obscured radio-loud AGN at $z\ge3$.   \citet{Middelberg2010} observed 17 IFRS and found that IFRS are predominantly steep spectrum objects with median spectral index $\alpha=-1.4$, consistent with high-redshift radio galaxies.  They reported polarization emission of three IFRS, one source having $\Pi_{\rm 0} = 12\,\%$ and the other two having $\Pi_{\rm 0}=7\,\%$.

\begin{figure}
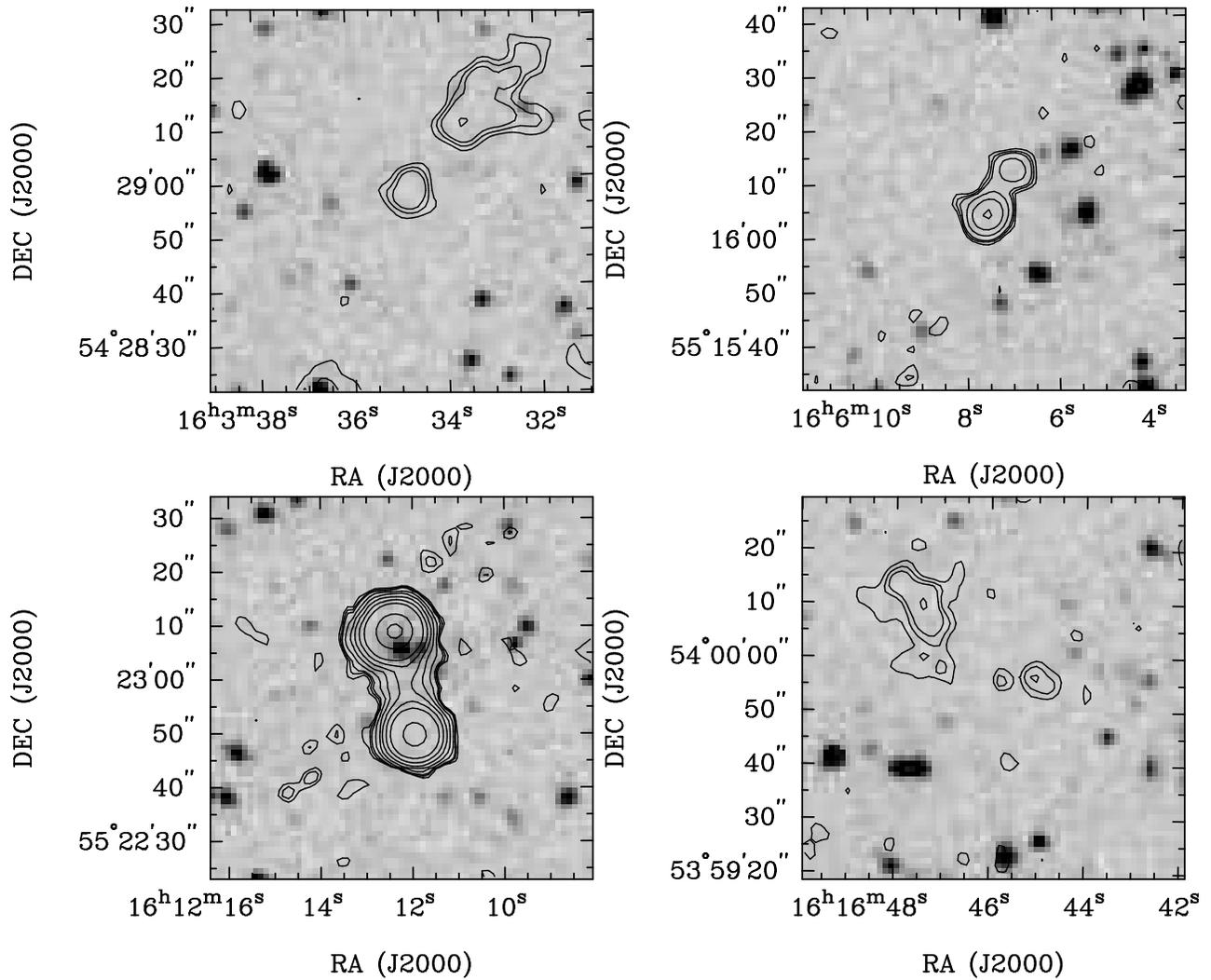

\includegraphics[scale=0.37,angle=-90]{655.swire.ps}
\includegraphics[scale=0.37,angle=-90]{494.swire.ps}
\includegraphics[scale=0.37,angle=-90]{1.swire.ps}
\includegraphics[scale=0.37,angle=-90]{693.swire.ps}
\caption{Polarized infrared faint radio sources with structure at arc-second scales.  The background image is the \textit{Spitzer} 3.6$\mu$m image overlaid with $1.4\,$GHz VLA contours starting at twice the local noise level and increase by a factor of 2 (2, 4, 6, 8, 16, 32, etc.).}
\label{ifrsimg}
\end{figure}

We found 18 IFRS in our sample of radio sources, of which five are polarized, see Table \ref{ifrstab}.  There are four polarized IFRS that show structure at arcsecond scales (Fig.$\,\,$\ref{ifrsimg}), while there are only two sources with no detectable polarization that show resolved structure.  The mean spectral index for our polarized IFRS is $-1.1$.  For the 13 sources with no detectable polarization the mean spectral index is $-1.5$.    The mean $\Pi_{\rm 0}$ of our polarized IFRS is $12.8\pm1.8\,\%$.  The flux densities of our polarized IFRS are $S_{\rm 1.4}< 9.0\,$mJy with one source having $S_{\rm 1.4}=360\,$mJy, while the IFRS with no detectable polarization have $S_{\rm 1.4}< 23\,$mJy.  Our spectral indices and percentage polarization are consistent with previous findings of \citet{Middelberg2010}.

\subsection{Polarization and Luminosity}\label{lumsec}
\begin{figure}
\begin{center}
\includegraphics[scale=0.75]{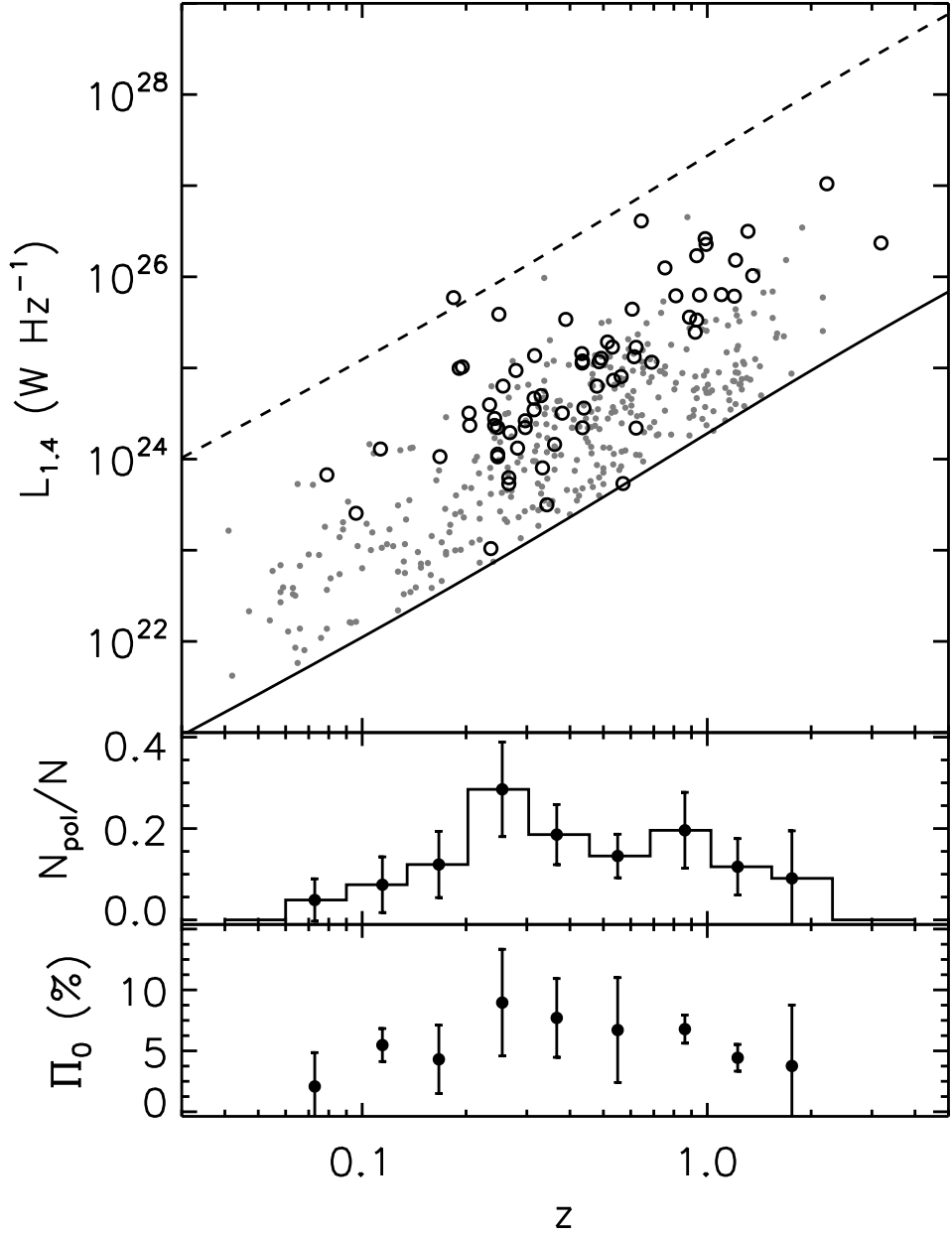}
\caption{\textit{Top}: Luminosity density and redshift relationship for polarized radio sources (open circles) in the DRAO deep field region covering the SWIRE observation area.  The grey dots are the sources with no detectable polarization.  The solid line indicates the flux density limit of our sample and the dashed line indicates the location of sources with a flux density $S_{\rm 1.4}=500\,$mJy. \textit{Middle}: The fraction of polarized sources compared to the total number of sources in the given redshift bins. \textit{Bottom}: The mean $\Pi_{\rm 0}$ of the polarized sources in the given redshift bins.} \label{lumred}
\end{center}
\end{figure}

We have examined luminosity density and redshift distributions of our polarized sample to study the effect of increasing $\Pi$ with decreasing flux density. Fig.$\,\,$\ref{lumred} shows the distribution of luminosity density as a function of redshift for both polarized radio sources and radio sources with no detectable polarization.  The range of redshifts in our sample is $0.04 < z_{\rm ph} < 3.2$.   As expected, the polarized sources are predominantly the more luminous radio galaxies, as these tend to be the higher flux density sources which have a low $\Pi$ detection threshold, a result of the detection limit in polarized flux density.  There appears to be an increase in the fraction of polarized sources with $0.2 < z < 0.3$.  

\begin{figure}
\begin{center}
\includegraphics[scale=0.8]{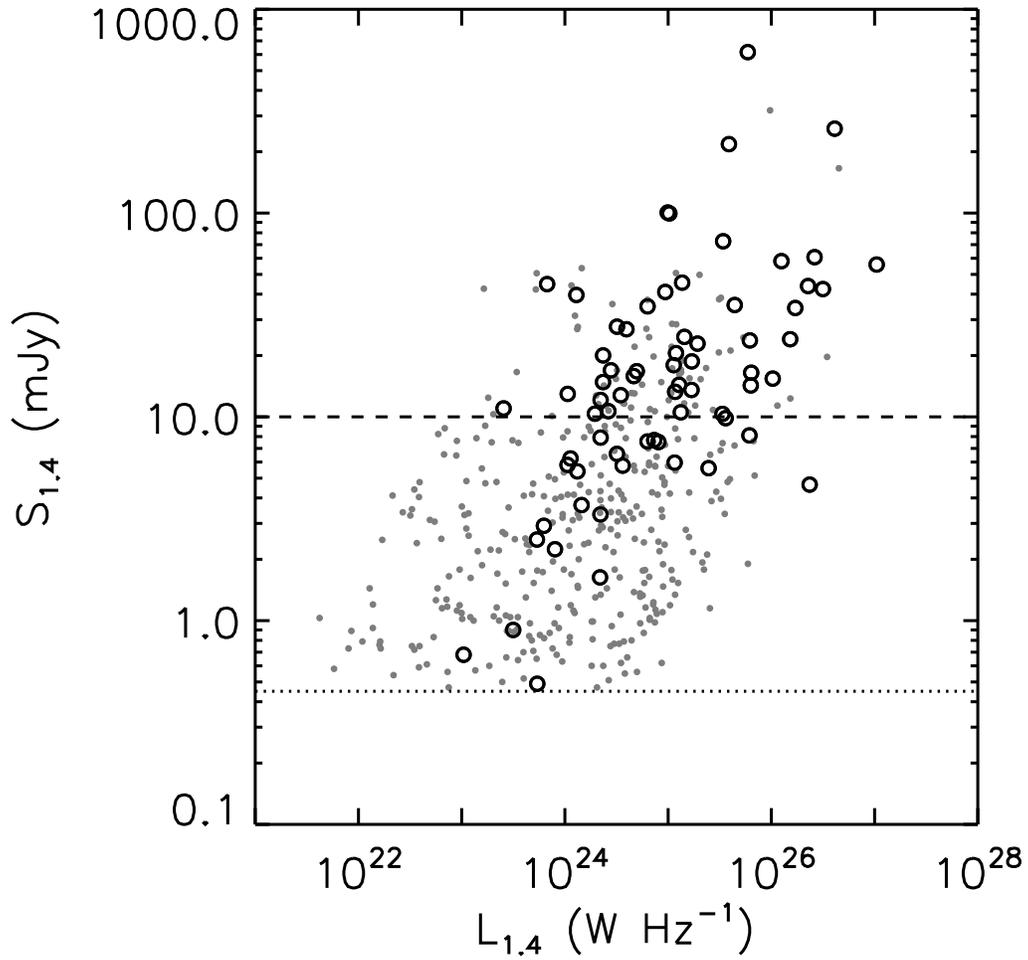}
\caption{Luminosity density and flux density at $1.4\,$GHz for all sources in the DRAO deep field.  The horizontal dotted line indicates the flux density detection limit and the horizontal dashed line indicates the flux density cut at $10\,$mJy. The open circles are polarized sources and the dots indicate the radio sources with no detectable polarization.} \label{lum_flux}
\end{center}
\end{figure}

\begin{figure}
\begin{center}
\includegraphics[scale=0.8]{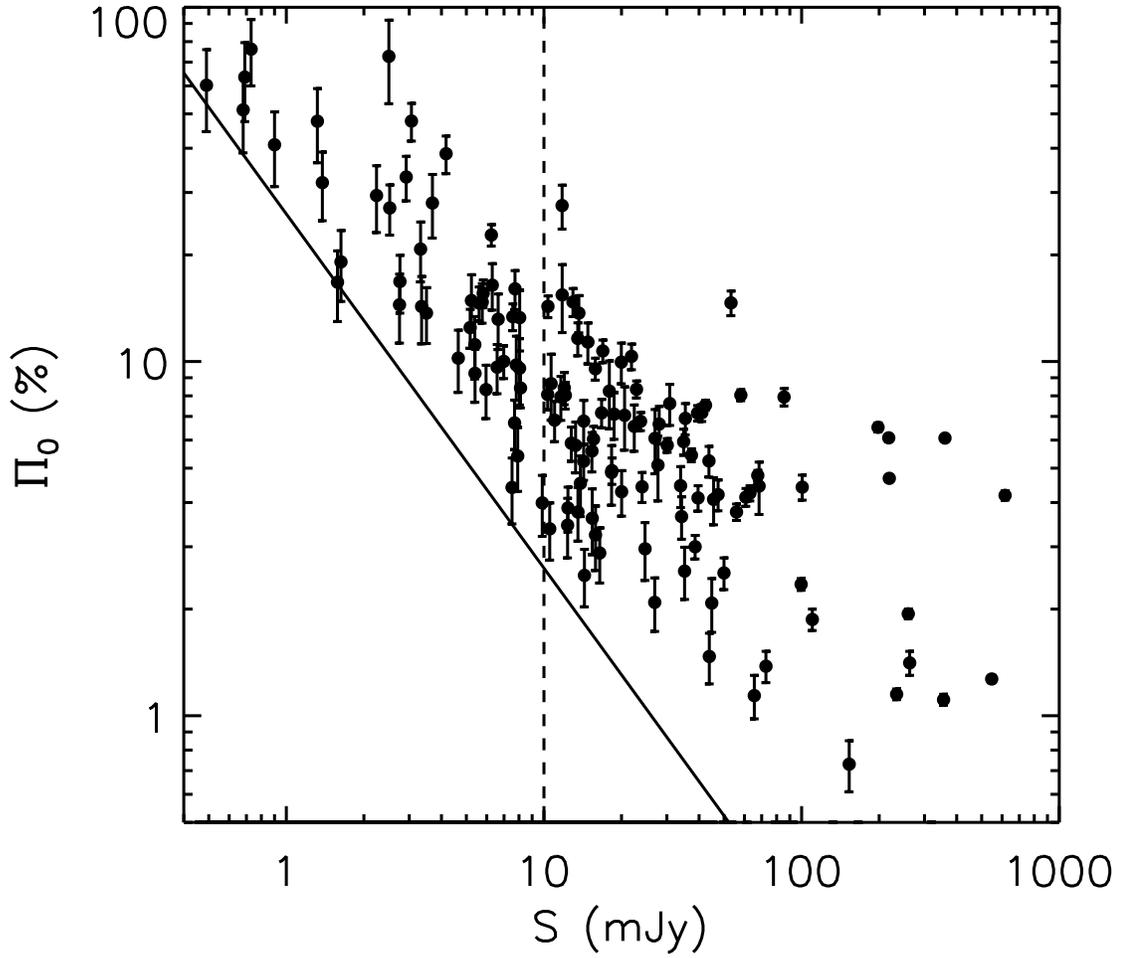}
\caption{$1.4\,$GHz flux density and bias-corrected percentage polarization for all polarized sources in the DRAO deep field.  The vertical dashed line is the $10\,$mJy flux density cut and the solid diagonal line is the percentage polarization detection limit.} \label{pi_flux}
\end{center}
\end{figure}

A complete flux limited sample is required to limit the selection effect in polarization.  Fig.$\,\,$\ref{lum_flux} plots luminosity density and flux density for all the radio sources in the DRAO deep field.  Applying a flux density cut of $S_{\rm 1.4}\ge10\,$mJy to our data, the polarized radio sources that remain in the analysis lie between $L_{\rm 1.4\,GHz} = 10^{23}$ and $10^{28}\,$W Hz$^{-1}$.  Fig.$\,\,$\ref{pi_flux} plots the bias-corrected percentage polarization ($\Pi_{\rm 0}$) and $1.4\,$GHz flux density for all polarized radio sources in the DRAO deep field. Our conservative flux density cutoff decreases the selection effect, i.e.~a polarized source can be detected at $\Pi_{\rm 0}=10\,\%$ at any flux density $S_{\rm 1.4} \ge 10\,$mJy.  We have applied the flux density cut of $S_{\rm 1.4}\ge 10\,$mJy for subsequent analysis.  

\begin{figure}
\begin{center}
\includegraphics[scale=0.5]{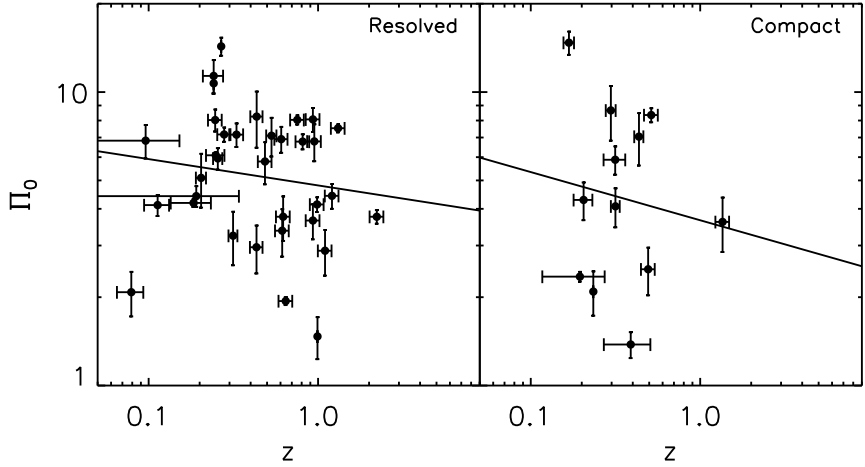}
\caption{Percentage polarization and redshift relationship for sources in the DRAO deep field polarized radio sample with $S_{\rm 1.4} \ge 10\,$mJy. \textit{Left}: The solid line is the best fit power-law with an index of $-0.09 \pm 0.11$ for resolved sources.  \textit{Right}:  The solid line is the best fit power-law with an index of $-0.16 \pm 0.38$ for compact sources.} \label{redpi}
\end{center}
\end{figure}

To study whether redshift contributes to this trend of increasing $\Pi_{\rm 0}$ with decreasing flux density, the polarized radio sources are plotted in Fig.$\,\,$\ref{redpi} as a function of percentage polarization and redshift.  \citet{Grant2010} split the polarized sources into compact and resolved classifications and suggest that the increase in $\Pi_{\rm 0}$ comes from the resolved sources.  Using the classification of compact and resolved sources from \citet{Grant2010}, we fit the data with a power-law of the form
\begin{equation}
\frac{\Pi\,\,}{\Pi_{\rm 0}} = \left(\frac{z}{z_{\rm 0}}\right)^\beta \ ,
\end{equation}
to the $12$ compact sources and $34$ resolved sources.  For resolved sources, $\beta=-0.09\,\pm\,0.11$ and for compact sources, $\beta=-0.16\,\pm\,0.38$.  The Pearson correlation coefficient was found to be $-0.14$ and $-0.13$ for resolved and compact sources respectively.  To measure the statistical dependence of $\Pi_{\rm 0}$ on redshift, the Spearman rank correlation coefficient was found to be $-0.17$ with a significance level of $0.33$ for resolved sources and $-0.11$ with significance level of $0.73$ for compact sources respectively.  The significance level ranges from 0 to 1 and smaller values indicates a significant correlation. Therefore, in our sample of polarized radio sources there is no correlation between redshift and percentage polarization.

\begin{figure}
\begin{center}
\includegraphics[scale=0.5]{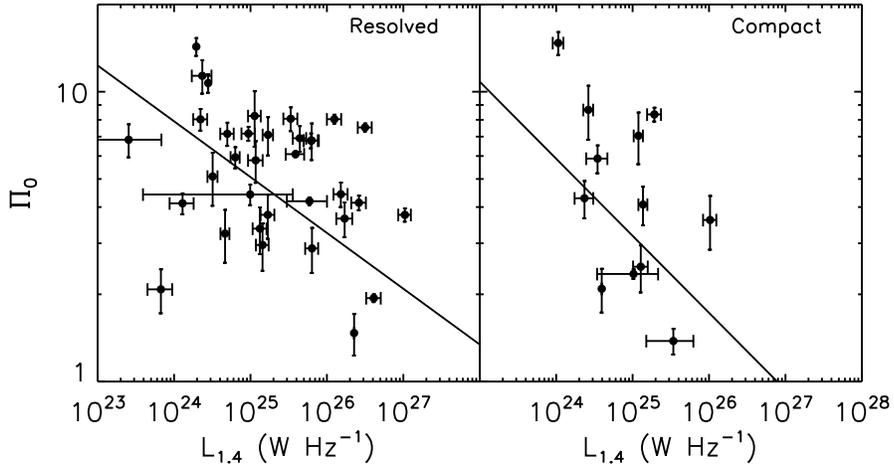}
\caption{\textit{Left}: Percentage polarization and luminosity relationship for resolved polarized sources in the DRAO deep field $S_{\rm 1.4} \ge 10\,$mJy.  The solid line is the best fit power-law with an index of $-0.19\pm0.06$.  \textit{Right}: Percentage polarization and luminosity relationship for compact polarized sources with $S_{\rm 1.4} \ge 10\,$mJy.  The solid line is the best fit power-law with an index of $-0.27\pm0.38$.} \label{lumpi_res}
\end{center}
\end{figure}

In Fig.$\,\,$\ref{lumpi_res} we plot luminosity density and percentage polarization for the 12 compact polarized radio sources and for the 34 resolved polarized radio sources with $S_{\rm 1.4} \ge 10\,$mJy.  A power-law of the form
\begin{equation}
\frac{\Pi\,\,}{\Pi_{\rm 0}} = \left(\frac{L_{\nu}}{L_{\rm 0}}\right)^\beta \ ,
\end{equation}
was fit for both the compact and resolved polarized radio sources separately.  The resolved sources were fit with $\beta=-0.19\,\pm\,0.06$, while the compact sources were fit with $\beta=-0.27\,\pm\,0.38$.  The Pearson correlation coefficient was found to be $-0.50$ and $-0.40$ for resolved and compact sources respectively.  To measure the statistical dependence of $\Pi_{\rm 0}$ on luminosity density, the Spearman rank correlation coefficient was found to be $-0.48$ with a significance level of $0.01$ for resolved sources and $-0.51$ with significance level of $0.03$ for compact sources respectively.  Therefore, $1.4\,$GHz observations show that resolved polarized radio sources exhibit a weak increase in $\Pi_{\rm 0}$ with decreasing luminosity.  The low-luminosity radio sources with low $\Pi_{\rm 0}$ are not detected in our sample, which is a result of our detection limit.

Summarizing, we found no evidence for a dependence of $\Pi_{\rm 0}$ on z, but we did find a weak dependence of $\Pi_{\rm 0}$ on luminosity density for polarized sources.  Sources resolved at $5\arcsec$ exhibit the trend of increasing $\Pi_{\rm 0}$ with decreasing luminosity density, while compact sources do not.

\section{Discussion}
Previous results have shown that $\Pi_{\rm 0}$ increases with decreasing flux density.  However, the reason for this effect remains unknown.  A systematic change in the properties of radio sources with decreasing flux density has been suggested by many authors as a possible explanation for this increase \citep[see][]{Mesa2002,Taylor2007,atlbs2010,Grant2010}.  We have compared both resolved sources and compact sources to study the properties of radio sources and the trend of increasing $\Pi_{\rm 0}$ with decreasing flux density.  The distribution of spectral index, percentage polarization, redshift, and luminosity density for the compact and resolved sources with a redshift and $S_{\rm 1.4}\ge 10\,$mJy are shown in Table \ref{radioprop}.  The compact and resolved sources differ in redshift and luminosity, with resolved sources spanning a larger range in both redshift and luminosity density.  However, the 12 compact radio sources showed no trend in $\Pi_{\rm 0}$ with redshift or luminosity density, while the 34 resolved radio sources show a weak trend with luminosity density but not with redshift.  To understand why the compact sources differ from the resolved sources, we have examined the possible AGN classifications of these objects.  

Our compact polarized radio sources could be BL Lac objects or Compact Steep Spectrum (CSS) radio sources.  \citet{Jackson1999} show that BL Lacertae objects are beamed flat-spectrum AGN of both FRI and FRII classification, and are known to have high and variable polarization.  \citet{Fan2006} studied 47 BL Lac objects and found that polarization correlates with spectral index.  \citet{Grant2010} showed that there is no such trend.  \citet{Saikia1999} studied the polarization properties of core-dominated quasars and BL Lac objects and found that the median $\Pi$ of these sources is similar at $3\,\%$.  The BL Lac objects and core-dominated quasars in the \citet{Saikia1999} sample were more highly polarized then the sample FRI and FRII radio sources.  

On the other hand, the polarized compact radio source may be CSS radio sources. CSS sources are radio sources that are unresolved at arc-second scales and are considered to be young radio sources with ages $\le 10^6\,$years \citep{Fanti1990}.  Polarization observations of CSS sources show that these objects are weakly polarized, with the median $\Pi_{\rm 0}$ of $2.4\,$\% at $1.4\,$GHz, but increasing with frequency \citep{Mantovani2009}.  The young ages of CSS sources imply that classical radio lobes have not yet formed and hence the radio emission probably originates in the central core of the host galaxy where there is a lot of dense and inhomogeneous gaseous material. Therefore the lower degree of percentage polarization may be caused by internal Faraday effects like the `partial coverage' of the Narrow-Line Region (NLR) as mentioned by \citet{Mantovani2009}.  The compact polarized radio sources in our sample can be a mixture of both types of radio sources.  The weak trend and poorer fit to compact sources for $\Pi_{\rm 0}$ and luminosity density suggests that these objects are a separate radio population from the resolved polarized sources and do not have dramatically evolving magnetic fields.  If, however, these compact objects are CSS sources and are the younger versions of the powerful radio-loud AGN, then this suggests there is magnetic field evolution from the compact radio sources to the resolved radio sources, since resolved radio sources tend to be more polarized \citep[see][]{Grant2010}.

\begin{figure}
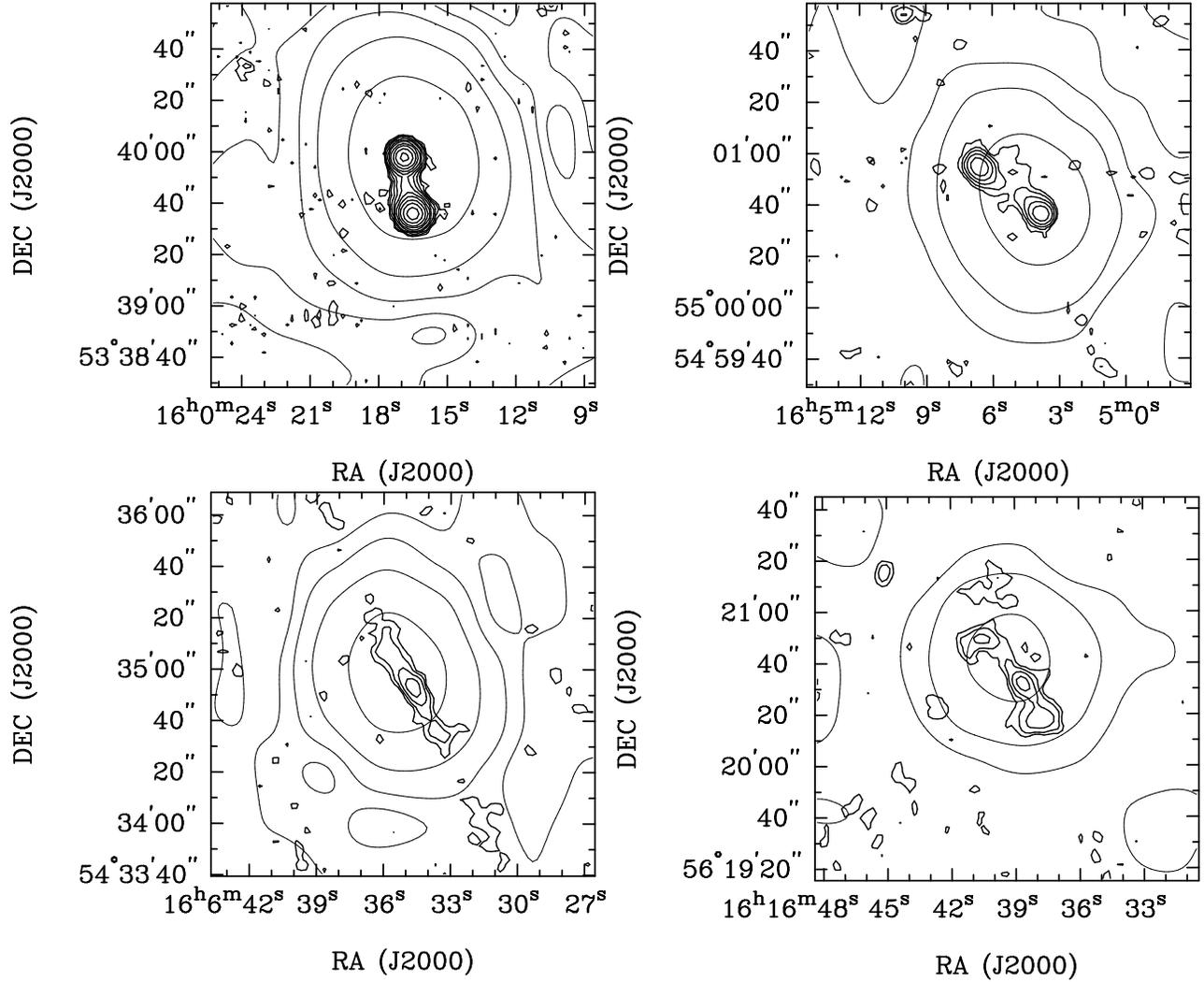

\includegraphics[scale=0.37,angle=-90]{20.first.ps}
\includegraphics[scale=0.37,angle=-90]{120.first.ps}
\includegraphics[scale=0.37,angle=-90]{171.first.ps}
\includegraphics[scale=0.37,angle=-90]{275.first.ps}
\caption{A selection of FRII (top row) and FRI (bottom row) polarized radio sources with a redshift.  The DRAO $1.4\,$GHz polarisation contours are represented by the grey lines and the VLA $1.4\,$GHz total intensity contours are shown in black.  The contours start at twice the local noise level and increase by a factor of 2 (2, 4, 6, 8, 16, 32, etc.) for both the DRAO and VLA data.}
\label{FRsample}
\end{figure}

\citet{Grant2010} showed that resolved radio sources are more highly polarized than compact radio sources.  The resolved polarized radio sources in our sample were found to have a weak trend of increasing $\Pi_{\rm 0}$ with decreasing luminosity density  but no trend with redshift.  The result of $\Pi_{\rm 0}$ not correlating with redshift is consistent with the study by \citet{Yang2001} who used $1518$ NVSS polarized radio sources with a redshift obtained from the catalog of Quasars and Active Nuclei by \citet{Veron2000}.  Visual inspection of the resolved radio sources show a mixture of FRIs and FRIIs.  Four resolved polarized radio sources are shown in Fig.$\,\,$ \ref{FRsample} to highlight the mixture of FRIs and FRIIs in our sample.  However, the location of the polarized emission is unknown.  

A study of $\Pi_{\rm 0}$ and luminosity density was carried out by \citet{Mesa2002} who found that there is no correlation between $\Pi_{\rm 0}$ and $L_{\rm 1.4}$ for sources with $S_{\rm 1.4} \ge 80\,$mJy.   \citet{Mesa2002} find a mean $\Pi_{\rm 0} = 2.9\,$\% for 143 steep-spectrum polarized radio sources at $1.4\,$GHz.  \citet{Goodlet2005} studied 26 powerful radio-loud AGNs and found a difference in RM in the separate lobes of the radio sources, indicating that the difference is internal to the source.  The difference in the RM of each lobe and the dispersion in RM both correlate with redshift but not with luminosity or source size, pointing to more turbulent environments for high redshift AGN compared to their low redshift counterparts.  In the cases of \citet{Mesa2002} and \citet{Goodlet2005}, the radio source samples are high flux density radio sources which are known to be less polarized than lower flux density radio sources.  Our study consists of polarized radio sources predominantly below $S_{\rm 1.4} = 80\,$mJy and our sources lie in the same luminosity range as \citet{Mesa2002}.  We do not find a trend of percentage polarization with redshift, indicating that our sources may be a different population than those studied by \citet{Mesa2002} and \citet{Goodlet2005}.  We are working on high resolution polarimetric observations which will provide the necessary information to determine the location of the polarized emission in both FRIs and FRIIs.  These higher resolution polarimetric observations will help to determine whether FRIs or FRIIs are more highly polarized.  However, we are aware that higher resolution imaging can potentially resolve out any diffuse polarized emission within the sources.  

\section{Conclusions}
We have presented polarization, redshift, luminosity and infrared classification of radio sources from the DRAO deep field source catalog.  There are 662 DRAO deep field radio sources with $S_{\rm 1.4} \ge 0.440\,$mJy in the SWIRE ELAIS N1 field of which 553 have high resolution $1.4\,$GHz positions from follow-up VLA observations.  The VLA observations cover the SWIRE ELAIS N1 region down to a rms noise level of $87\,\mu$Jy beam$^{-1}$, a factor of 2 deeper than FIRST.  Of the 553 radio sources with high resolution positions, 444 have an unambiguous match to a SWIRE object.  

The infrared fluxes from SWIRE show that the majority of polarized radio sources in our sample have hosts that are old elliptical galaxies with an imbedded AGN.  These galaxies can be either high or low luminosity radio sources.  The rest of the polarized radio sources have host galaxies where the infrared emission originates from light reprocessed by dust and gas and where the radio emission comes from the AGN.  The AGN hosted by elliptical galaxies are more highly polarized than the AGN with infrared emission from dust and gas.  Correlating these polarized radio galaxies with redshifts show no trend between $\Pi_{\rm 0}$ and redshift.  

We were nevertheless able to determine a weak correlation between $\Pi_{\rm 0}$ and luminosity.    The trend of increasing $\Pi_{\rm 0}$ with decreasing $L_{\rm 1.4}$ shows that low-luminosity radio sources are more polarized than high-luminosity radio sources for sources in the flux density range of  $S_{\rm 1.4} = 10 - 80\,$mJy.  There is a mixture of polarized FRIs, FRIIs and compact sources in the two different types of host galaxy, therefore we are unable to determine if FRIs are more polarized than FRIIs.  High resolution polarization imaging will be able to determine the origin of polarized emission in these sources and whether FRIs or FRIIs are more polarized.

\acknowledgements
We thank R.~Perley, M.~Goss, and M.~Claussen for their support and discussions with VLA image processing.  J.~K.~B. acknowledges support from the Natural Sciences and Engineering Research Council of Canada.   Observations and research on the DRAO Planck Deep Fields are supported by the Natural Science and Engineering Council of Canada and the National Research Council Canada. The Dominion Radio Astrophysical Observatory is operated as a National Facility by the National Research Council of Canada.  This research has made use of observations made with the \textit{Spitzer Space Telescope} and the NASA/IPAC Extragalactic Database (NED) which both are operated by the Jet Propulsion Laboratory, California Institute of Technology under a contract with NASA.  Funding for the SDSS and SDSS-II has been provided by the Alfred P. Sloan Foundation, the Participating Institutions, the National Science Foundation, the U.S. Department of Energy, the National Aeronautics and Space Administration, the Japanese Monbukagakusho, the Max Planck Society, and the Higher Education Funding Council for England. The SDSS Web Site is http://www.sdss.org/.

\clearpage

%%% TABLES %%%
%%%%%%%%%%%

% Available ELAIS N1 observations
\begin{deluxetable}{cccc}
\tablewidth{0pt}
\tablecaption{Available data and observations in the ELAIS N1 region.}
\tablehead{
  \colhead{Wavelength} & \colhead{Instrument} & \colhead{Sensitivity (5$\sigma$)} & \colhead{Reference}
}
\startdata
3.6, 4.5, 5.8, $8.0\,\mu$m  & SWIRE$-$IRAC & 7.3, 9.7, 27.5, $32.5\,\mu$Jy & \citet{Lonsdale2003}\\
\\
24, 70, $160\,\mu$m & SWIRE$-$MIPS & $450\,\mu$Jy, 2.75, $17.5\,$mJy & \citet{Lonsdale2003}\\
\\
7,15, 90, $175\mu$m & {\it ISO} & $5\,$mJy & \citet{Oliver2000}\\
\\
$0.5-8\,$keV & {\it Chandra} & $75\,$ks & \citet{Manners2003}\\
\\
$u'$,$g'$,$r'$,$i'$,$z'$ & INT$-$WFS & 23.4, 24.9, 24.0, 23.2, $21.9\,$mag & \citet{McMahon2001}\\
\\
$21\,$cm ($1.4\,$GHz) & VLA & $135\,\mu$Jy & \citet{Ciliegi1999}\\
 & & $435\,\mu$Jy & this paper\\
 & VLA$-$FIRST & $1\,$mJy & \citet{White1997}\\
 & VLA$-$NVSS & $2\,$mJy & \citet{Condon1998}\\
 & DRAO ST & $400\,\mu$Jy & \citet{Taylor2007}\\
 & DRAO ST & $390\,\mu$Jy, polarization & \citet{Taylor2007}\\ 
  & DRAO ST & $275\,\mu$Jy & \citet{Grant2010}\\
 & DRAO ST & $225\,\mu$Jy, polarization & \citet{Grant2010}\\
 \\
$49\,$cm ($610\,$MHz) & GMRT & $200\,\mu$Jy & \citet{Garn2008}\\
\\
$92\,$cm ($325\,$MHz) & Westerbork ST & $18\,$mJy & \citet{Rengelink1997}\\
 & GMRT & $200\,\mu$Jy & \citet{Sirothia2009}\\
\enddata
\label{enobs}
\end{deluxetable}

% classification of SWIRE matches
\begin{deluxetable}{lrr}
\tablewidth{0pt}
\tablecaption{Breakdown of DRAO deep field radio sources located in the SWIRE ELAIS N1 observations}
\tablehead{
  \colhead{Cross-Identification} & \colhead{Polarized Sources}  & \colhead{No Polarization}
}
\startdata
Unambiguous  & 70 (65$\pm$8\%) & 374 (67$\pm$3\%)\\
Ambiguous & 21 (20$\pm$4\%) & 145 (26$\pm$2\%)\\
Infrared Faint Radio Sources & 5 (5$\pm$2\%) & 13 (3$\pm$1\%)\\
Not cataloged & 11 (10$\pm$3\%) & 23 (4$\pm$1\%)\\ 
\enddata
\label{swireid}
\end{deluxetable}

% merged source catalog
\begin{deluxetable}{lrrrrccc}
\tablewidth{0pt}
\tabletypesize{\scriptsize}
\tablecaption{Polarized sources from the DRAO deep field radio source catalog which have a redshift.  The spectral indices are calculated from the $1.4\,$GHz and $325\,$MHz flux density from WENSS, $p_{\rm 0}$ and $\Pi_{\rm 0}$ are the bias-corrected polarized flux density and percentage polarization.  The flux density and polarization information comes from the \citet{Grant2010} source catalog.}
\tablehead{
 \colhead{DRAO} & \colhead{SWIRE} & \colhead{$S_{\rm 1.4}$} & \colhead{$p_{\rm 0}$} & \colhead{$\Pi_{\rm 0}$} & \colhead{$\alpha^{1420}_{325}$}& \colhead{$z_{\rm ph}$} & \colhead{$L_{\rm 1.4}$}\\
 \colhead{ } & \colhead{ } & \colhead{(mJy)} & \colhead{(mJy)} & \colhead{(\%)} & \colhead{ }& \colhead{ } & \colhead{(W Hz$^{-1}$)}
}
\startdata
DRAOJ\_160016.66+533944.82&\nodata&$616.34\pm  8.58$&$ 25.81\pm  0.74$&$  4.19\pm  0.13$&$ -1.01\pm  0.07$&$0.184^b$&$ 5.909\times25$\\
DRAOJ\_160505.45+550045.50&    374150&$ 34.21\pm  0.72$&$  1.25\pm  0.17$&$  3.65\pm  0.50$&$ -1.19\pm  0.08$&$0.932^a$&$ 1.705\times26$\\
DRAOJ\_160600.00+545406.16&    374873&$259.46\pm  3.18$&$  5.04\pm  0.14$&$  1.94\pm  0.06$&$ -0.77\pm  0.07$&$0.644^a$&$ 4.105\times26$\\
DRAOJ\_160634.97+543455.67&    365142&$ 16.92\pm  0.42$&$  1.81\pm  0.13$&$ 10.71\pm  0.79$&$ -0.69\pm  0.17$&$0.242^c$&$ 2.794\times24$\\
DRAOJ\_160721.60+534639.58&    151535&$ 15.40\pm  0.49$&$  0.56\pm  0.12$&$  3.61\pm  0.76$&$ -0.41\pm  0.27$&$1.355^a$&$ 1.031\times26$\\
DRAOJ\_160722.80+553059.65&    408285&$ 14.26\pm  0.48$&$  0.97\pm  0.14$&$  6.79\pm  0.98$&$ -0.94\pm  0.15$&$0.950^a$&$ 6.315\times25$\\
DRAOJ\_160724.12+531408.34&\nodata&$ 10.66\pm  0.49$&$  0.92\pm  0.19$&$  8.66\pm  1.83$&$ -0.49\pm  0.35$&$0.297^b$&$ 2.637\times24$\\
DRAOJ\_160828.34+541031.91&    175922&$ 26.91\pm  0.52$&$  0.56\pm  0.10$&$  2.09\pm  0.36$&$ -0.47\pm  0.15$&$0.234^c$&$ 3.946\times24$\\
DRAOJ\_160830.91+534421.98&    156843&$ 13.27\pm  0.45$&$  0.77\pm  0.12$&$  5.80\pm  0.95$&$ -0.95\pm  0.16$&$0.486^a$&$ 1.169\times25$\\
DRAOJ\_160907.44+535425.60&    168323&$ 43.84\pm  0.74$&$  0.64\pm  0.11$&$  1.47\pm  0.24$&$ -1.01\pm  0.08$&$0.992^c$&$ 2.268\times26$\\
DRAOJ\_160910.22+552632.14&    415349&$  5.76\pm  0.24$&$  0.85\pm  0.10$&$ 14.71\pm  1.88$&$ -0.69\pm  0.47$&$0.439^a$&$ 3.628\times24$\\
DRAOJ\_160922.66+561511.20&\nodata&$ 35.38\pm  0.84$&$  2.44\pm  0.24$&$  6.91\pm  0.70$&$ -0.57\pm  0.11$&$0.606^b$&$ 4.415\times25$\\
DRAOJ\_160944.14+543749.37&    202770&$  9.85\pm  0.30$&$  0.39\pm  0.08$&$  3.99\pm  0.78$&$> -0.41$&$0.888^a$&$ 3.607\times25$\\
DRAOJ\_160953.30+550708.29&    405410&$  0.90\pm  0.11$&$  0.37\pm  0.07$&$ 40.92\pm  9.74$&$> -2.03$&$0.343^a$&$ 3.155\times23$\\
DRAOJ\_161002.52+541637.63&    190161&$ 16.46\pm  0.36$&$  0.47\pm  0.08$&$  2.88\pm  0.51$&$ -0.28\pm  0.30$&$1.099^a$&$ 6.378\times25$\\
DRAOJ\_161012.89+544925.90&    213445&$  1.63\pm  0.12$&$  0.31\pm  0.07$&$ 19.11\pm  4.33$&$> -1.63$&$0.622^a$&$ 2.190\times24$\\
DRAOJ\_161027.53+541246.94&    189954&$ 10.36\pm  0.32$&$  1.48\pm  0.09$&$ 14.31\pm  1.01$&$ -0.29\pm  0.46$&$0.268^c$&$ 1.953\times24$\\
DRAOJ\_161037.54+532430.78&\nodata&$ 44.81\pm  0.84$&$  0.93\pm  0.16$&$  2.08\pm  0.36$&$ -0.63\pm  0.09$&$0.079^b$&$ 6.726\times23$\\
DRAOJ\_161057.53+553528.00&    431174&$ 22.87\pm  0.45$&$  1.91\pm  0.10$&$  8.35\pm  0.46$&$ -0.52\pm  0.16$&$0.514^a$&$ 1.926\times25$\\
DRAOJ\_161119.85+552846.31&    428862&$ 23.71\pm  0.45$&$  1.61\pm  0.09$&$  6.78\pm  0.40$&$ -0.70\pm  0.13$&$0.811^a$&$ 6.191\times25$\\
DRAOJ\_161138.18+535924.86&    187983&$ 24.04\pm  0.47$&$  1.07\pm  0.10$&$  4.43\pm  0.43$&$ -0.65\pm  0.14$&$1.208^a$&$ 1.522\times26$\\
DRAOJ\_161149.32+550049.82&    231824&$  5.81\pm  0.20$&$  0.91\pm  0.08$&$ 15.59\pm  1.41$&$ -0.93\pm  0.33$&$0.247^a$&$ 1.060\times24$\\
DRAOJ\_161159.98+555720.23&\nodata&$ 15.85\pm  0.42$&$  0.51\pm  0.11$&$  3.24\pm  0.67$&$ -0.64\pm  0.19$&$0.315^b$&$ 4.648\times24$\\
DRAOJ\_161223.71+552554.91&\nodata&$ 12.77\pm  0.30$&$  0.75\pm  0.08$&$  5.88\pm  0.65$&$ -0.37\pm  0.34$&$0.315^b$&$ 3.478\times24$\\
DRAOJ\_161229.02+550636.72&    239645&$ 14.36\pm  0.31$&$  0.36\pm  0.07$&$  2.49\pm  0.46$&$ -0.89\pm  0.16$&$0.493^a$&$ 1.278\times25$\\
DRAOJ\_161249.32+550239.05&    238910&$ 10.53\pm  0.26$&$  0.36\pm  0.06$&$  3.37\pm  0.62$&$ -0.53\pm  0.33$&$0.614^a$&$ 1.328\times25$\\
DRAOJ\_161250.59+560358.07&\nodata&$  3.32\pm  0.22$&$  0.69\pm  0.12$&$ 20.77\pm  3.99$&$> -1.15$&$0.435^b$&$ 2.209\times24$\\
DRAOJ\_161303.24+543224.47&    218306&$ 10.34\pm  0.26$&$  0.83\pm  0.08$&$  8.07\pm  0.76$&$ -0.54\pm  0.33$&$0.932^a$&$ 3.358\times25$\\
DRAOJ\_161320.50+541636.80&    209853&$  6.25\pm  0.22$&$  1.42\pm  0.09$&$ 22.76\pm  1.59$&$> -0.72$&$0.247^a$&$ 1.133\times24$\\
DRAOJ\_161331.25+542718.14&\nodata&$ 99.58\pm  1.35$&$  2.34\pm  0.08$&$  2.35\pm  0.09$&$ -0.70\pm  0.07$&$0.195^b$&$ 1.028\times25$\\
DRAOJ\_161341.76+561215.37&\nodata&$ 12.96\pm  0.44$&$  1.91\pm  0.16$&$ 14.73\pm  1.34$&$ -1.31\pm  0.12$&$0.168^b$&$ 1.064\times24$\\
DRAOJ\_161355.42+545753.14&    241939&$  0.49\pm  0.08$&$  0.29\pm  0.06$&$ 60.25\pm 15.69$&$> -2.45$&$0.570^a$&$ 5.396\times23$\\
DRAOJ\_161400.34+535715.08&     46196&$ 16.74\pm  0.39$&$  1.20\pm  0.11$&$  7.16\pm  0.66$&$ -0.32\pm  0.28$&$0.330^a$&$ 4.971\times24$\\
DRAOJ\_161412.14+554132.75&    454481&$  5.40\pm  0.22$&$  0.50\pm  0.08$&$  9.26\pm  1.59$&$> -0.82$&$0.282^a$&$ 1.321\times24$\\
DRAOJ\_161421.62+553651.26&    451943&$ 42.37\pm  0.67$&$  3.19\pm  0.09$&$  7.53\pm  0.24$&$ -0.62\pm  0.10$&$1.312^a$&$ 3.157\times26$\\
DRAOJ\_161509.94+531015.82&\nodata&$100.57\pm  1.79$&$  4.44\pm  0.36$&$  4.42\pm  0.36$&$ -0.69\pm  0.07$&$0.191^b$&$ 9.906\times24$\\
DRAOJ\_161527.67+542712.46&    230443&$  7.57\pm  0.26$&$  1.01\pm  0.08$&$ 13.38\pm  1.16$&$> -0.59$&$0.479^a$&$ 6.320\times24$\\
DRAOJ\_161537.82+534646.96&     48475&$ 60.76\pm  0.99$&$  2.51\pm  0.14$&$  4.14\pm  0.24$&$ -0.77\pm  0.08$&$0.986^a$&$ 2.622\times26$\\
DRAOJ\_161547.14+532820.21&\nodata&$  3.69\pm  0.30$&$  1.03\pm  0.19$&$ 28.02\pm  5.71$&$> -1.08$&$0.361^b$&$ 1.450\times24$\\
DRAOJ\_161558.82+552505.48&    271343&$  0.68\pm  0.10$&$  0.35\pm  0.07$&$ 51.34\pm 12.52$&$> -2.22$&$0.236^a$&$ 1.044\times23$\\
DRAOJ\_161559.21+532416.56&\nodata&$217.89\pm  3.04$&$ 13.26\pm  0.24$&$  6.09\pm  0.14$&$ -0.74\pm  0.07$&$0.249^b$&$ 3.881\times25$\\
DRAOJ\_161603.82+540434.46&     60000&$  4.65\pm  0.22$&$  0.48\pm  0.09$&$ 10.22\pm  2.04$&$> -0.92$&$3.188^a$&$ 2.350\times26$\\
DRAOJ\_161616.49+541724.65&     68227&$  5.96\pm  0.23$&$  0.50\pm  0.08$&$  8.33\pm  1.43$&$ -0.85\pm  0.36$&$0.690^a$&$ 1.156\times25$\\
DRAOJ\_161623.93+552703.82&    274949&$ 12.10\pm  0.29$&$  0.97\pm  0.08$&$  8.03\pm  0.69$&$ -0.95\pm  0.17$&$0.247^a$&$ 2.218\times24$\\
DRAOJ\_161639.14+562032.64&\nodata&$ 18.71\pm  0.63$&$  1.33\pm  0.20$&$  7.10\pm  1.07$&$ -0.52\pm  0.20$&$0.531^b$&$ 1.695\times25$\\
DRAOJ\_161639.46+554525.74&\nodata&$ 72.72\pm  1.63$&$  1.00\pm  0.10$&$  1.38\pm  0.14$&$ -0.63\pm  0.08$&$0.389^b$&$ 3.411\times25$\\
DRAOJ\_161758.06+543518.64&     85013&$  7.52\pm  0.26$&$  0.33\pm  0.07$&$  4.41\pm  0.93$&$> -0.59$&$0.563^a$&$ 8.051\times24$\\
DRAOJ\_161807.58+544237.87&     89241&$  8.11\pm  0.27$&$  0.68\pm  0.08$&$  8.42\pm  1.02$&$> -0.54$&$1.198^a$&$ 6.133\times25$\\
DRAOJ\_161826.28+542532.77&     82324&$ 13.55\pm  0.33$&$  0.51\pm  0.09$&$  3.76\pm  0.65$&$ -0.44\pm  0.29$&$0.622^a$&$ 1.685\times25$\\
DRAOJ\_161831.99+543834.58&     88998&$  7.69\pm  0.26$&$  0.52\pm  0.08$&$  6.71\pm  1.07$&$> -0.58$&$0.535^a$&$ 7.332\times24$\\
DRAOJ\_161833.00+543145.77&     85921&$ 58.09\pm  1.81$&$  4.67\pm  0.09$&$  8.04\pm  0.29$&$ -0.66\pm  0.09$&$0.754^a$&$ 1.252\times26$\\
DRAOJ\_161915.31+550513.27&    278015&$  5.60\pm  0.21$&$  0.82\pm  0.08$&$ 14.62\pm  1.61$&$ -1.04\pm  0.29$&$0.923^a$&$ 2.468\times25$\\
DRAOJ\_161919.97+553600.97&    298450&$ 55.89\pm  1.17$&$  2.10\pm  0.10$&$  3.76\pm  0.20$&$ -0.41\pm  0.10$&$2.221^a$&$ 1.048\times27$\\
DRAOJ\_161923.38+540032.98&\nodata&$ 41.01\pm  0.76$&$  2.94\pm  0.15$&$  7.17\pm  0.40$&$ -0.73\pm  0.09$&$0.279^b$&$ 9.389\times24$\\
DRAOJ\_161924.79+555113.57&\nodata&$  2.92\pm  0.20$&$  0.97\pm  0.12$&$ 33.18\pm  4.77$&$> -1.23$&$0.266^b$&$ 6.256\times23$\\
DRAOJ\_162006.53+543233.11&     93142&$  6.58\pm  0.25$&$  0.63\pm  0.10$&$  9.64\pm  1.53$&$> -0.68$&$0.380^a$&$ 3.194\times24$\\
DRAOJ\_162011.28+562615.43&\nodata&$ 20.55\pm  0.78$&$  1.45\pm  0.29$&$  7.05\pm  1.43$&$ -0.52\pm  0.19$&$0.435^b$&$ 1.192\times25$\\
DRAOJ\_162038.02+545129.41&\nodata&$ 11.00\pm  0.34$&$  0.75\pm  0.10$&$  6.83\pm  0.89$&$ -0.92\pm  0.19$&$0.096^b$&$ 2.547\times23$\\
DRAOJ\_162047.54+545037.25&\nodata&$  7.91\pm  0.28$&$  0.43\pm  0.09$&$  5.41\pm  1.11$&$ -0.97\pm  0.24$&$0.297^b$&$ 2.217\times24$\\
DRAOJ\_162145.50+542111.34&\nodata&$  2.24\pm  0.20$&$  0.66\pm  0.13$&$ 29.42\pm  6.29$&$> -1.41$&$0.333^b$&$ 8.013\times23$\\
DRAOJ\_162145.98+554950.20&\nodata&$ 34.85\pm  0.72$&$  2.07\pm  0.17$&$  5.94\pm  0.50$&$ -0.55\pm  0.12$&$0.256^b$&$ 6.322\times24$\\
DRAOJ\_162209.53+552341.46&\nodata&$ 39.63\pm  0.71$&$  1.63\pm  0.13$&$  4.12\pm  0.34$&$ -0.87\pm  0.09$&$0.113^b$&$ 1.291\times24$\\
DRAOJ\_162216.46+552209.59&\nodata&$ 20.02\pm  0.47$&$  0.86\pm  0.12$&$  4.29\pm  0.63$&$ -0.79\pm  0.14$&$0.205^b$&$ 2.344\times24$\\
DRAOJ\_162233.58+553035.14&\nodata&$ 24.68\pm  0.56$&$  0.73\pm  0.13$&$  2.96\pm  0.55$&$ -0.56\pm  0.15$&$0.433^b$&$ 1.438\times25$\\
DRAOJ\_162321.72+541418.78&\nodata&$ 14.81\pm  0.54$&$  1.68\pm  0.21$&$ 11.35\pm  1.50$&$ -0.48\pm  0.26$&$0.242^b$&$ 2.341\times24$\\
DRAOJ\_162500.26+555020.58&\nodata&$ 27.70\pm  0.94$&$  1.41\pm  0.29$&$  5.10\pm  1.06$&$ -0.77\pm  0.12$&$0.204^b$&$ 3.197\times24$\\
DRAOJ\_162513.99+560032.00&\nodata&$  2.50\pm  0.40$&$  1.82\pm  0.38$&$ 72.69\pm 19.27$&$> -1.34$&$0.266^b$&$ 5.356\times23$\\
DRAOJ\_162602.54+553858.16&\nodata&$ 17.94\pm  0.77$&$  1.48\pm  0.31$&$  8.25\pm  1.79$&$ -0.77\pm  0.16$&$0.434^b$&$ 1.133\times25$\\
DRAOJ\_162607.27+550300.76&\nodata&$ 45.55\pm  1.03$&$  1.86\pm  0.28$&$  4.08\pm  0.62$&$ -0.70\pm  0.09$&$0.316^b$&$ 1.367\times25$\\
\enddata
\tablenotetext{a}{Photometric redshift from \citet{RR2008}.  $^{\rm b}$ Photometric redshift from SDSS DR7 \citep{sdssdr7}. $^{\rm c}$ Spectroscopic redshift from \citet{RR2008} and references within.}
\label{catalog}
\end{deluxetable}

\begin{deluxetable}{lrr}
\tablewidth{0pt}
\tablecaption{DRAO deep field radio sources in the SWIRE color-color diagram.}
\tablehead{
\colhead{Region} & \colhead{Polarized Sources}  & \colhead{Other Sources}
}
\startdata
1 &  12 (27$\,\pm\,$8\%)& 42 (21$\,\pm\,$3\%) \\
2 &  2 (4$\,\pm\,$3\%) & 28 (14$\,\pm\,$3\%) \\
3a & 20 (45$\,\pm\,$10\%) & 87 (44$\,\pm\,$5\%) \\
3b & 10 (22$\,\pm\,$7\%) & 32 (16$\,\pm\,$3\%) \\
4 & 1 (2$\,\pm\,$2\%) & 9 (5$\,\pm\,$2\%) \\
\enddata
\label{swirenum}
\end{deluxetable}

%%%
%  Spiral Galaxy table
%%%
\begin{deluxetable}{lllrrr}
\tablewidth{0pt}
\tabletypesize{\footnotesize}
\tablecaption{DRAO deep field radio sources matched to spiral galaxies in the SWIRE ELAIS N1 field with upper limits on percentage polarization $\Pi_{\rm UL}$.}
\tablehead{
  \colhead{DRAO ID} & \colhead{Name} & \colhead{Type}  & \colhead{$z$} & \colhead{$S_{1.4}$} & \colhead{$\Pi_{\rm UL}$} \\
  \colhead{ } & \colhead{ } & \colhead{ }  & \colhead{ } & \colhead{(mJy)} & \colhead{(\%)}
}
\startdata
DRAO\_J160603.24+552527.62 & UGC 10214 & SB(s)c pec & $0.06^1$ & $1.57\,\pm\,0.12$ & $<\,$29\\
DRAO\_J160736.67+535729.92 & CGCG 275$-$024 & Sbc & $0.03^2$ & $1.24\,\pm\,0.04$ & $<\,$21 \\
DRAO\_J161035.26+561613.40 &  CGCG 275$-$026 & \nodata & 0.06$^3$ & $4.40\,\pm\,0.06$ & $<\,$14 \\
DRAO\_J161109.62+535811.96 & MCG +09$-$26$-$063 & Sy 2 & 0.10$^1$ & $0.89\,\pm\,0.06$ & $<\,$31 \\
DRAO\_J161237.03+535814.02 & MCG +09$-$26$-$066 & \nodata & 0.07$^3$ & $1.30\,\pm\,0.08$ & $<\,$21 \\
DRAO\_J161254.58+545524.74 & 2MASX J16125415+5455261 & Sa & 0.12$^1$ & $0.58\,\pm\,0.08$ & $<\,$29 \\
DRAO\_J161331.51+541637.34 & 2MASX J16133121+5416294 & S0/a & 0.08$^1$ & $1.44\,\pm\,0.15$ & $<\,$17 \\
DRAO\_J161449.82+554509.11 & 2MASX J16144902+5545120 & \nodata & 0.09$^3$ & $0.76\,\pm\,0.06$ & $<\,$36 \\
DRAO\_J161457.41+555224.71 &  CGCG 276$-$004 & \nodata & 0.08$^3$ & $6.46\,\pm\,0.55$ & $<\,$4 \\
DRAO\_J161459.30+535507.43 &  SDSS J161459.13+535507.4 & \nodata & 0.15$^1$ & $1.15\,\pm\,0.05$ & $<\,$27 \\
DRAO\_J161819.44+541901.34 & 2MASX J16181934+5418587 & \nodata & 0.13$^1$ & $1.44\,\pm\,0.08$ & $<\,$11 \\
DRAO\_J161905.64+550252.40 & 2MASX J16190575+5502447 & \nodata & 0.16$^1$ & $0.59\,\pm\,0.06$ & $<\,$27 \\
DRAO\_J162142.36+550507.04 & NGC 6143 & SAB(rs)bc & 0.06$^3$ & $4.78\,\pm\,0.06$ & $<\,$5 \\
DRAO\_J162144.90+542725.92 & SBS 1620+545 & Sy 2 & 0.10$^3$ & $3.63\,\pm\,0.06$ & $<\,$7 \\
DRAO\_J162148.05+543927.94 & MCG +09$-$27$-$023 & \nodata & 0.10$^3$ & $6.41\,\pm\,0.13$ & $<\,$4\\
DRAO\_J162211.14+550255.90 & MCG +09$-$27$-$025 & \nodata & 0.08$^3$ & $3.12\,\pm\,0.07$ & $<\,$8\\
\enddata
\tablenotetext{1}{Photometric redshift from \cite{RR2008}. $^2$Spectroscopic redshift and $^3$Photometric redshift from SDSS DR7 \citep{sdssdr7}.}
\label{spiral}
\end{deluxetable}

\begin{deluxetable}{lrrrr}
%\rotate
\tablewidth{0pt}
\tabletypesize{\footnotesize}
\tablecaption{DRAO deep field infrared faint radio sources. The spectral indices are calculated from the $1.4\,$GHz and $325\,$MHz flux density from WENSS, $p_{\rm 0}$ and $\Pi_{\rm 0}$ are the bias-corrected polarized flux density and percentage polarization.  The flux density and polarization information comes from the \citet{Grant2010} source catalog.}
\tablehead{
  \colhead{DRAO ID} & \colhead{$S_{\rm 1.4}$} & \colhead{$p_{\rm0}$}  & \colhead{$\Pi_{\rm 0}$} & \colhead{$\alpha^{1420}_{325}$} \\
  \colhead{ } & \colhead{(mJy)}  & \colhead{(mJy)} & \colhead{(\%)} & \colhead{}
}
\startdata
DRAOJ\_160333.60+542906.79 &    8.05 $\pm$ 0.41 & 1.07 $\pm$ 0.20 & 13.30 $\pm$ 2.59 & $-$1.07 $\pm$ 0.21 \\
DRAOJ\_160422.03+550546.07 &   22.22 $\pm$ 0.69 & \nodata         & \nodata & $-$1.26 $\pm$ 0.09 \\
DRAOJ\_160530.48+540902.09 &    3.56 $\pm$ 0.24 & \nodata         & \nodata & $>\,-$1.10 \\
DRAOJ\_160552.97+551032.52 &    3.12 $\pm$ 0.23 & \nodata         & \nodata & $-$1.29 $\pm$ 0.37 \\
DRAOJ\_160607.30+551608.04 &    6.30 $\pm$ 0.31 & 1.04 $\pm$ 0.15 & 16.43 $\pm$ 2.46 & $-$0.54 $\pm$ 0.53 \\
DRAOJ\_160647.93+541510.51 &   2.23 $\pm$ 0.18 & \nodata         & \nodata & $>\,-$1.42 \\
DRAOJ\_160838.74+542751.88 &   1.39 $\pm$ 0.13 & \nodata         & \nodata & $>\,-$1.73 \\
DRAOJ\_160949.75+540833.32 &    1.08 $\pm$ 0.13 & \nodata         & \nodata & $>\,-$1.91 \\
DRAOJ\_161112.89+543317.64 &    2.45 $\pm$ 0.14 & \nodata         & \nodata & $>\,-$1.35 \\
DRAOJ\_161212.29+552302.18 &  360.15 $\pm$ 4.20 & 21.90 $\pm$ 0.07 & 6.08 $\pm$ 0.07 & $-$1.01 $\pm$ 0.07 \\
DRAOJ\_161225.78+545503.04 &    0.84 $\pm$ 0.10 & \nodata         & \nodata & $>\,-$2.08 \\
DRAOJ\_161324.86+553930.64 &    3.50 $\pm$ 0.18 & 0.48 $\pm$ 0.08 & 13.70 $\pm$ 2.46 & $>\,-$1.11 \\
DRAOJ\_161647.42+535951.00 &  3.35 $\pm$ 0.20 & 0.48 $\pm$ 0.10 & 14.30 $\pm$ 3.08 & $-$1.83 $\pm$ 0.19 \\
DRAOJ\_161818.38+554310.60 &   1.90 $\pm$ 0.15 & \nodata         & \nodata & $>\,-$1.52 \\
DRAOJ\_161832.09+545105.44 &    0.66 $\pm$ 0.10 & \nodata         & \nodata & $>\,-$2.24 \\
DRAOJ\_162033.94+544328.85 &   1.72 $\pm$ 0.14 & \nodata         & \nodata & $>\,-$1.59 \\
DRAOJ\_162153.35+545321.77 &   6.52 $\pm$ 0.27 & \nodata         & \nodata & $-$0.94 $\pm$ 0.30 \\
DRAOJ\_162408.74+545218.73 &    6.96 $\pm$ 0.35 & \nodata         & \nodata & $>\,-$0.64 \\
\enddata
\label{ifrstab}
\end{deluxetable}

\begin{deluxetable}{lccccc}
%\rotate
\tablewidth{0pt}
\tabletypesize{\footnotesize}
\tablecaption{Radio source properties for the resolved and compact polarized sources in our sample with a redshift and $S_{\rm 1.4} \ge 10\,$mJy.  Spectral index, $\Pi_{\rm 0}$, redshift, and luminosity density are presented as the range of values with the minimum value first.}
\tablehead{
  \colhead{ } & \colhead{N} & \colhead{$\alpha^{1420}_{325}$}  & \colhead{$\Pi_{\rm 0}$} & \colhead{$z$} & \colhead{$L_{\rm 1.4}$} \\
  \colhead{ } & \colhead{ } &  \colhead{ } & \colhead{(\%)} & \colhead{} & \colhead{(W Hz$^{-1}$)}
}
\startdata
Resolves Sources & 34 & $-1.19$, $-0.28$ & $1.5$, $14.3$ & $0.08$, $2.22$ & $2.55\times 10^{23}$, $1.04 \times 10^{27}$\\
Compact Sources & 12 & $-1.31$, $-0.37$ & $1.4$, $14.8$ & $0.17$, $1.36$ & $1.06\times 10^{24}$, $1.03\times 10^{26}$\\
\enddata
\label{radioprop}
\end{deluxetable}
%%%%%%%%
%                   %
%  FIGURES %
%                   %
%%%%%%%%
%%%
%  VLA observing area
%%%

%%%SDSS-zph plot

%%% color-color plot

%%% luminosity plots

\end{document}